\documentclass[aps,prc,superscriptaddress,showpacs,amssymb,amsmath,amsfonts,preprint,floatfix]{revtex4}
\usepackage{graphicx}

\begin{document}

\newcommand{\piplusRxn}{\gamma \ p \rightarrow n \ \pi^+}
\newcommand{\xRxn}{\gamma \ p \rightarrow \pi^+ X}
\newcommand{\xRxnp}{\gamma \ p \rightarrow \pi^+ X}
\newcommand{\ppipiRxn}{\gamma \ p \rightarrow p \ \pi^- \ \pi^+}
\newcommand{\ppipiyRxn}{\gamma \ p \rightarrow p \ \pi^- \ \pi^+ \ Y}
\newcommand{\Eg}{E_{\gamma}}
\newcommand{\cosThetaCm}{\cos\theta_{\rm c.m.}}
\newcommand{\cosThetaCmP}{\cos\theta^p_{\rm c.m.}}
\newcommand{\cosThetaCmPi}{\cos\theta^{\pi}_{\rm c.m.}}

\title{$\pi^+$ photoproduction on the proton for photon energies from
0.725 to 2.875~GeV}

\newcommand*{\ANL}{Argonne National Laboratory}
\newcommand*{\ANLindex}{1}
\affiliation{\ANL}
\newcommand*{\ASU}{Arizona State University, Tempe, Arizona 85287-1504}
\newcommand*{\ASUindex}{2}
\affiliation{\ASU}
\newcommand*{\CSU}{California State University, Dominguez Hills, Carson, CA 90747}
\newcommand*{\CSUindex}{3}
\affiliation{\CSU}
\newcommand*{\CMU}{Carnegie Mellon University, Pittsburgh, Pennsylvania 15213}
\newcommand*{\CMUindex}{4}
\affiliation{\CMU}
\newcommand*{\CUA}{Catholic University of America, Washington, D.C. 20064}
\newcommand*{\CUAindex}{5}
\affiliation{\CUA}
\newcommand*{\SACLAY}{CEA-Saclay, Service de Physique Nucl\'eaire, 91191 Gif-sur-Yvette, France}
\newcommand*{\SACLAYindex}{6}
\affiliation{\SACLAY}
\newcommand*{\CNU}{Christopher Newport University, Newport News, Virginia 23606}
\newcommand*{\CNUindex}{7}
\affiliation{\CNU}
\newcommand*{\UCONN}{University of Connecticut, Storrs, Connecticut 06269}
\newcommand*{\UCONNindex}{8}
\affiliation{\UCONN}
\newcommand*{\ECOSSEE}{Edinburgh University, Edinburgh EH9 3JZ, United Kingdom}
\newcommand*{\ECOSSEEindex}{9}
\affiliation{\ECOSSEE}
\newcommand*{\FU}{Fairfield University, Fairfield CT 06824}
\newcommand*{\FUindex}{10}
\affiliation{\FU}
\newcommand*{\FIU}{Florida International University, Miami, Florida 33199}
\newcommand*{\FIUindex}{11}
\affiliation{\FIU}
\newcommand*{\FSU}{Florida State University, Tallahassee, Florida 32306}
\newcommand*{\FSUindex}{12}
\affiliation{\FSU}
\newcommand*{\GWU}{The George Washington University, Washington, DC 20052}
\newcommand*{\GWUindex}{13}
\affiliation{\GWU}
\newcommand*{\ECOSSEG}{University of Glasgow, Glasgow G12 8QQ, United Kingdom}
\newcommand*{\ECOSSEGindex}{14}
\affiliation{\ECOSSEG}
\newcommand*{\ISU}{Idaho State University, Pocatello, Idaho 83209}
\newcommand*{\ISUindex}{15}
\affiliation{\ISU}
\newcommand*{\INFNFR}{INFN, Laboratori Nazionali di Frascati, 00044 Frascati, Italy}
\newcommand*{\INFNFRindex}{16}
\affiliation{\INFNFR}
\newcommand*{\INFNGE}{INFN, Sezione di Genova, 16146 Genova, Italy}
\newcommand*{\INFNGEindex}{17}
\affiliation{\INFNGE}
\newcommand*{\ORSAY}{Institut de Physique Nucleaire ORSAY, Orsay, France}
\newcommand*{\ORSAYindex}{18}
\affiliation{\ORSAY}
\newcommand*{\ITEP}{Institute of Theoretical and Experimental Physics, Moscow, 117259, Russia}
\newcommand*{\ITEPindex}{19}
\affiliation{\ITEP}
\newcommand*{\JMU}{James Madison University, Harrisonburg, Virginia 22807}
\newcommand*{\JMUindex}{20}
\affiliation{\JMU}
\newcommand*{\KYUNGPOOK}{Kyungpook National University, Daegu 702-701, Republic of Korea}
\newcommand*{\KYUNGPOOKindex}{21}
\affiliation{\KYUNGPOOK}
\newcommand*{\UNH}{University of New Hampshire, Durham, New Hampshire 03824-3568}
\newcommand*{\UNHindex}{22}
\affiliation{\UNH}
\newcommand*{\NSU}{Norfolk State University, Norfolk, Virginia 23504}
\newcommand*{\NSUindex}{23}
\affiliation{\NSU}
\newcommand*{\OHIOU}{Ohio University, Athens, Ohio  45701}
\newcommand*{\OHIOUindex}{24}
\affiliation{\OHIOU}
\newcommand*{\ODU}{Old Dominion University, Norfolk, Virginia 23529}
\newcommand*{\ODUindex}{25}
\affiliation{\ODU}
\newcommand*{\RPI}{Rensselaer Polytechnic Institute, Troy, New York 12180-3590}
\newcommand*{\RPIindex}{26}
\affiliation{\RPI}
\newcommand*{\URICH}{University of Richmond, Richmond, Virginia 23173}
\newcommand*{\URICHindex}{27}
\affiliation{\URICH}
\newcommand*{\MOSCOW}{Skobeltsyn Nuclear Physics Institute, Skobeltsyn Nuclear Physics Institute, 119899 Moscow, Russia}
\newcommand*{\MOSCOWindex}{28}
\affiliation{\MOSCOW}
\newcommand*{\SCAROLINA}{University of South Carolina, Columbia, South Carolina 29208}
\newcommand*{\SCAROLINAindex}{29}
\affiliation{\SCAROLINA}
\newcommand*{\JLAB}{Thomas Jefferson National Accelerator Facility, Newport News, Virginia 23606}
\newcommand*{\JLABindex}{30}
\affiliation{\JLAB}
\newcommand*{\UNIONC}{Union College, Schenectady, NY 12308}
\newcommand*{\UNIONCindex}{31}
\affiliation{\UNIONC}
\newcommand*{\UTFSM}{Universidad T\'{e}cnica Federico Santa Mar\'{i}a, Casilla 110-V Valpara\'{i}so, Chile}
\newcommand*{\UTFSMindex}{32}
\affiliation{\UTFSM}
\newcommand*{\VT}{Virginia Polytechnic Institute and State University, Blacksburg, Virginia   24061-0435}
\newcommand*{\VTindex}{33}
\affiliation{\VT}
\newcommand*{\VIRGINIA}{University of Virginia, Charlottesville, Virginia 22901}
\newcommand*{\VIRGINIAindex}{34}
\affiliation{\VIRGINIA}
\newcommand*{\WM}{College of William and Mary, Williamsburg, Virginia 23187-8795}
\newcommand*{\WMindex}{35}
\affiliation{\WM}
\newcommand*{\YEREVAN}{Yerevan Physics Institute, 375036 Yerevan, Armenia}
\newcommand*{\YEREVANindex}{36}
\affiliation{\YEREVAN}

\newcommand*{\NOWVIRGINIA}{University of Virginia, Charlottesville, Virginia 22901}
\newcommand*{\NOWJLAB}{Thomas Jefferson National Accelerator Facility, Newport News, Virginia 23606}
\newcommand*{\NOWUTFSM}{Universidad T\'{e}cnica Federico Santa Mar\'{i}a, Casilla 110-V Valpara\'{i}so, Chile}
\newcommand*{\NOWGWU}{The George Washington University, Washington, DC 20052}
\newcommand*{\NOWECOSSEE}{Edinburgh University, Edinburgh EH9 3JZ, United Kingdom}
\author {M.~Dugger}
\affiliation{\ASU}
\author {B.G.~Ritchie}
\affiliation{\ASU}
\author {J.P.~Ball}
\affiliation{\ASU}
\author {P.~Collins}
\affiliation{\ASU}
\author {E.~Pasyuk}
\affiliation{\ASU}
\author {R.A. Arndt}
\affiliation{\GWU}
\author {W.J.~Briscoe}
\affiliation{\GWU}
\author {I.I.~Strakovsky}
\affiliation{\GWU}
\author {R.L.~Workman}
\affiliation{\GWU}
\author {M.J.~Amaryan} 
\affiliation{\ODU}
% ...

\collaboration{The CLAS Collaboration}
\noaffiliation

\begin{abstract}
Differential cross sections for the reaction $\piplusRxn$ have been
measured with the CEBAF Large Acceptance Spectrometer (CLAS) and a
tagged photon beam with energies from 0.725 to 2.875~GeV.  Where
available, the results obtained here compare well with previously
published results for the reaction.  Agreement with the SAID and MAID
analyses is found below 1~GeV. The present set of cross sections has been
incorporated into the SAID database, and exploratory fits have been
made up to 2.7~GeV. Resonance couplings have been extracted and
compared to previous determinations. With the addition of these cross
sections to the world data set, significant changes have occurred in
the high-energy behavior of the SAID cross-section predictions and
amplitudes.
\end{abstract}

\maketitle

\section{Introduction}
\label{sec:Intro}

The photoproduction of mesons has played a crucial role in the search
for resonances beyond those found through analyses of pion-nucleon
elastic scattering data. Cross section structures seen in kaon and eta
photoproduction~\cite{etc} have been interpreted as candidates for
so-called ``missing'' resonances, excitations that are predicted by
QCD-inspired models~\cite{caro} but expected to couple weakly to the
pion-nucleon channel.

The photoproduction of pions, though less likely to detect states not
seen in pion-nucleon studies, is the most well-developed of the
meson-photoproduction programs, having an extensive database for which
many single- and multi-channel fits are available~\cite{GW}. The
photo-decay amplitudes for non-strange resonances have been determined
almost exclusively from this reaction~\cite{PDG}.  However, while
cross section data exist, they are quite sparse above an incident
photon energy $E_\gamma =$1.7~GeV, and have generally come from
untagged bremsstrahlung measurements. As a result, all photo-decay
amplitudes for the higher $N^\ast$ states have an inherent uncertainty
beyond any model-dependence due to the background-resonance extraction
process. While some theory-based model dependence is unavoidable,
cross sections measured precisely using a tagged-photon beam, with
incident photon energies covering the full resonance region, will
provide tighter and more reliable constraints for future analyses of
the properties of excited nucleons.

In this paper, we report measurements of the unpolarized differential
cross sections for $\pi^+$ photoproduction on the proton for
$E_\gamma$ from 0.725 to 2.875~GeV. As a first step to gauge their
influence, we have included these new cross sections in a multipole
fit to all available data covering the resonance region. This task is
aided by the inclusion of tagged neutral pion cross sections recently
measured~\cite{dugger} that span a range in $E_\gamma$ from 0.675 to
2.875~GeV.  We have obtained a revised set of multipole amplitudes and
have extracted photo-decay couplings for those states that couple
strongly to the pion-nucleon final state. Using the revised multipole
analysis, we have generated predictions for further measurements of
polarization observables that should soon become available.

The paper is laid out in the following manner: We give a brief
background of the experimental parameters for this study in
Sec.~\ref{sec:Exp}.  An overview of the method used is given in
Sec.~\ref{sec:Data}. The uncertainty estimates for the cross sections
obtained are given in Sec.~\ref{sec:Errs}.  The experimental results
are described in Sec.~\ref{sec:Results}. Various fits to the data are
described in Sec.~\ref{sec:fit}, and the underlying multipole
amplitudes and resonance contributions are displayed and compared to
previous determinations in Sec.~\ref{sec:ResCoupl}.  Finally, in
Sec.~\ref{sec:conc}, we provide a brief summary of the results of the
study and consider what extensions of this work would be particularly
helpful in the future.

\section{Experiment}
\label{sec:Exp}


The differential cross sections for the reaction $\piplusRxn$ were
measured with the CEBAF Large Acceptance Spectrometer
(CLAS)~\cite{CLAS} and the bremsstrahlung photon-tagging facility
(``photon tagger'')~\cite{tag} in Hall B of the Thomas Jefferson
National Accelerator Facility (JLab) as part of a set of experiments
running at the same time with the same experimental configuration
(cryogenic target, tagger, and CLAS) called the ``{\tt{g1c}}'' run
period.  The cross sections were part of a program of meson
photoproduction measurements undertaken using CLAS and the photon
tagger~\cite{dugger,meson1,meson2,meson3,meson4,meson5,meson6,meson7,meson8}.

The data described here were obtained in sets of data runs with differing 
energies for the electron beam incident on the photon tagger.
The two incident electron
energies were 2.445 and 3.115~GeV. Moreover, the 3.115~GeV data runs were
taken with either the full photon-tagger focal plane (``3.115-full'')
or higher-photon-energy half of the
photon-tagger focal plane (``3.115-half'') in operation. Thus, 
for example, during the 3.115-half
running, data was accumulated only for the higher-energy half of the available
photon energies in order to increase statistics for data collected at those 
higher energies.  

The produced tagged photons impinged on an
18-cm-long liquid-hydrogen target placed at the center of CLAS.  This
target was enclosed by a scintillator array (called the ``start
counter,'' described in Ref.~\cite{start}) that detected the passage
of charged particles into CLAS from the target.  The event trigger
required the coincidence of a post-bremsstrahlung electron passing
through the focal plane of the photon tagger and at least one charged
particle detected in CLAS and the start counter.  Tracking of the
charged particles through the magnetic field within CLAS by drift
chambers \cite{DC} provided determination of their charge, momentum,
and scattering angle.  This information, together with the particle
velocity measured by the time-of-flight system \cite{TOF} and start
counter, provided particle identification for each particle detected
in CLAS and its corresponding momentum four-vector.  

The methods used
for extracting the differential cross sections for $\pi^+$
photoproduction are presented in the next several sections.  The
technique is outlined initially, and then each step is described in
further detail, with a summary provided of the data and tests that support the
validity of the approach taken.

\section{Data Reduction}
\label{sec:Data}

The technique for this analysis is very similar to that used
previously in the analysis of the CLAS {\tt{g1c}} running period data
for the reaction $\gamma p \rightarrow p \pi^0$ \cite{dugger}.  In
that analysis, the recoiling proton from the photoproduction process
was detected in CLAS and, assuming the two-body reaction $\gamma \ p
\rightarrow p X$ (where $X$ was the undetected $\pi^0$), yields were
determined in the missing mass spectra for the reconstructed $\pi^0$.

In this analysis, similarly, the photoproduced $\pi^+$ was detected in
CLAS.  Again assuming the two-body reaction $\xRxn$, where in the
present case $X$ was the undetected neutron, yields were determined in
the missing mass spectra for the reconstructed neutron.  However,
while both the proton and $\pi^+$ are positively charged particles,
the CLAS detector response to the recoiling pions and protons was
different (for example, the amount of energy deposited in the
scintillators within the detector), which necessitated appropriate
modifications to the previous analysis.

For the data described in this paper, yields for the neutron were
determined using the following steps:

\newcounter{bean}
\begin{list}{\arabic{bean}.}{\usecounter{bean}\setlength{\rightmargin}{\leftmargin}}
\item Identify the $\pi^+$ in CLAS, determining the scattering angle and
momentum.
\item Sort the events in the resulting missing mass spectra into kinematic
bins in incident photon energy $\Eg$ and scattering angle
$\theta^{\pi}_{\rm c.m.}$, where $\theta^{\pi}_{\rm c.m.}$ is the
center-of-mass angle of the $\pi^+$.
\item Identify the missing mass peak for the neutron in each kinematic bin.
\item Determine the yield for the neutron in each kinematic bin by
subtracting the background beneath the peak.
\item Correct the meson yield in each kinematic bin for spectrometer
acceptance using a Monte Carlo simulation of the spectrometer
acceptance.
\item Normalize the measured yield in each kinematic bin using a measured
absolute photon flux normalization procedure, thereby determining the
differential cross section for that bin.
\end{list}

In the following sections, each of these steps is described.  Also
presented are sample results, and, in some cases, tests that establish
the validity of the procedures used.

\subsection{Particle identification and kinematic variables}
\label{sec:Event}


\begin{figure}
\includegraphics[scale=0.4]{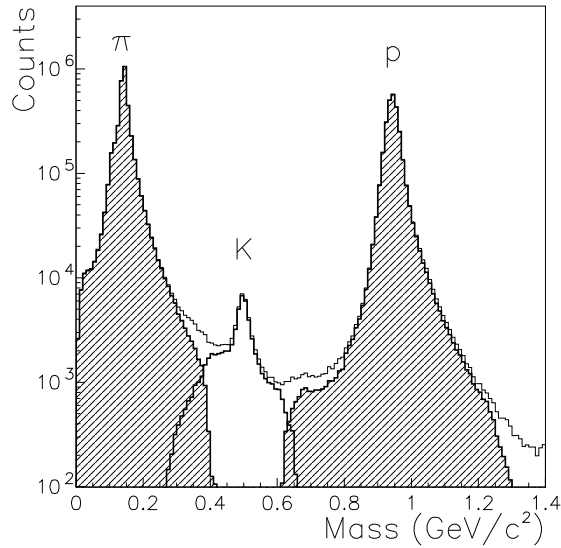}
\caption{ Particle identification spectrum obtained with CLAS, showing
identifications provided by the GPID algorithm (discussed in the text)
for all charged
particles.
\label{fig:pid}}
\end{figure}

The tracking information provided by the drift chambers within CLAS
gave momentum and scattering angle information on charged particles
scattered within the detector volume.  Time-of-flight and start
counter information, coupled with the track information provided by
the drift chambers, determined particle velocity and momentum.

Particle identification in this analysis was performed using the GPID
algorithm \cite{gpid}.  The method uses the momentum of the detected
particle, and sequentially calculates trial values of the velocity
$\beta$ for all possible particle identities.  Each one of the
possible identities is tested by comparing the trial value of $\beta$
for a given particle type to the empirically measured value of $\beta$
(as determined by CLAS tracking and time-of-flight information).  The
particle is assigned the identity that provides the closest trial
value of $\beta$ to the empirically measured value of $\beta$.  For
example, if the curvature indicates a positive particle, the $\beta$
is calculated for $p$, $\pi^+$, and $K^+$.  Figure\ \ref{fig:pid}
shows the mass distribution of the identified charged particles.  The
GPID algorithm also attempts to find a matching photon in the tagging
system for every charged particle detected in CLAS.  A matched photon
means that there was one and only one tagged photon in the trigger
window, which, in this analysis, was 18 ns.  Particles that were
determined not to have a matching photon are considered to be a
measure of the accidentals (to be described in more detail in the next
subsection).

CLAS is divided into six sectors in azimuthal angle.  Geometrical
fiducial cuts in each of the six sectors of CLAS were imposed on all
pions.  The region selected for accepting pions in each sector
corresponded to a region of relatively uniform detection efficiency
(constant to $\pm$ 3\%) versus azimuthal angle.

\subsection{Missing mass reconstruction}

\begin{figure}
\includegraphics[scale=0.4]{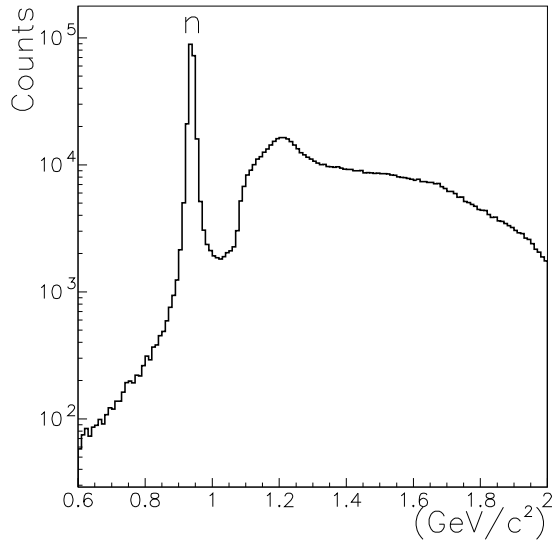}
\caption{ Missing mass spectrum obtained from the {\tt{g1c}} data set using
CLAS, assuming the reaction \protect{$\xRxn$} .}
\label{fig:mm}
\end{figure}

The momentum for the $\pi^+$ was determined by the drift chamber
system.  The momentum determined by CLAS was corrected for energy loss
in both the target cell and the start counter \cite{eloss}.  The
scattering angle and momentum was used to calculate the missing mass
based on the assumption that the reaction observed is $\pi^+ X$.
Based on this assumption, the missing mass spectrum in the full
spectrometer acceptance for all photon energies is shown in Fig.\
\ref{fig:mm}. The neutron peak is clearly seen.

Taking each $\pi^+$ event that did not have a matching incident photon
as noted above, and integrating over all of the out-of-time (not
within the trigger coincidence window) incident photons for that
event, determined the distribution of accidental coincidences between
CLAS and the photon tagger. This assumes that coupling the out-of-time
tagger hits to unmatched pion created a fair representation of the
accidental coincidences between CLAS and tagger.

\subsection{Distribution of events into kinematic bins}

The events from both the 2.445 and 3.115 GeV data sets, constituting
the full missing mass spectrum described in the previous section, were
sorted into bins in incident photon energy $\Eg$ and $\cosThetaCmPi$.
The widths of these ``kinematic bins'' ($\Delta\Eg = 50$ MeV in photon
energy and $\Delta\cosThetaCmPi$ = 0.1) were chosen such that, in
general, there were at least 1000 $\pi^+ n$ events in each kinematic
bin.


\subsection{\label{sec:Yield}Neutron yield}

For each kinematic bin, the neutron yield was extracted by removing
the background under the peak.  We have proceeded with the assumption
that the background in the missing mass spectra arises from two
particular types of events: \newcounter{bean2}
\begin{list}{\arabic{bean2}.}{\usecounter{bean2}\setlength{\rightmargin}{\leftmargin}}
\item Events arising from accidental coincidences between CLAS and the
photon tagger, as discussed in the preceding subsection.
\item Events arising from two-pion photoproduction via the reaction
$\xRxn$, where $X = p \pi^-$ or $X = n \pi^0$.
\end{list}

\begin{figure}
\includegraphics[scale=0.4]{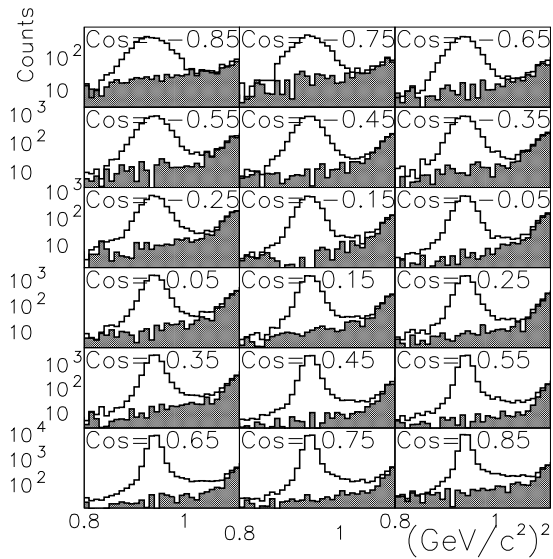}
\caption{ Neutron yield extraction for $\Eg = 1.475$ GeV.  The
background is represented as the shaded region.
\label{fig:fit1450}}
\end{figure}

The spectrum for accidental coincidences is determined by looking at
events that fell outside the designated trigger window.  To determine
the two-pion background, data for the reaction $\ppipiRxn$ were
selected by requiring that each particle in the final state \ had to
be identified through normal particle identification procedures, that
the same incident photon was chosen for each particle, and that the
missing mass was consistent with zero.  These selected data were used
to determine the {\em{shape}} of the $X = p \pi^-$ and $X = n \pi^0$
components of the background from two-pion photoproduction (due to
$\Delta(1232)$ dominance, the contribution from the $X = n \pi^0$
reaction was assumed to have the same shape as the $X = p \pi^-$
contribution).  This shape was used to generate the background beneath
the neutron peak, which was then subtracted from the neutron yield for
each kinematic bin. 
The fractional uncertainty in the background beneath the peak was
statistically added in quadrature to the uncertainty in the yield for each
kinematic bin. In most cases (>93 \%), the peak-to-background ratio was
greater than 5 to 1; in all cases, the signal-to-background ratio was
greater than 1.4 to 1.
Fig.\ \ref{fig:fit1450} shows an example of this background removal
procedure for all kinematic bins with photon energy $E_{\gamma} =
1.475$ GeV.

\subsection{Acceptance and efficiency}
\label{sec:Acceptance}


The spectrometer acceptance for charged pions was determined from the
results of Monte Carlo simulations of the CLAS detector response to
positive pions.  As a preliminary test of the quality of the Monte
Carlo representation of the CLAS response to $\pi^+$, simulated
acceptances for $\pi^+$ were compared to empirical measurements based
on the reaction $\gamma p \rightarrow p \pi^- \pi^+$ for most of the
kinematic bins in this study (the empirical acceptance method is not
useful for some regions of phase space due to limited statistics for
those kinematic bins).  Such an empirical check is practical for
much of the phase space covered in this experiment due to the large
number of events for that final state, and that all of the final
products leave charged tracks in CLAS, making them easily observable.
For the empirical comparison, in addition to the $\pi^+$, the proton
and $\pi^-$ were required to be detected in the event and both were
assigned the same photon.  The same fiducial cuts applied to the
$\pi^+$ noted above were applied to both reconstructed and
CLAS-identified $\pi^+$.  A missing mass reconstruction from the
kinematic information of the proton and $\pi^-$ was performed to
determine if a $\pi^+$ should have been seen in CLAS.  The background
beneath this peak was removed by subtracting a polynomial fit (order
3) to the background region from the spectrum.

A comparison of Monte-Carlo-simulated events to actual data for the
$\ppipiRxn$ reaction (re-binned as if the $\pi^+$ came from the
$\piplusRxn$ reaction channel) was performed.  Simulated events were
obtained by generating $10^7$ $\ppipiRxn$ events that were
isotropic in phase space and then processed through a GEANT simulation
of CLAS created by the Jefferson Laboratory {\tt{GSIM}} working group.
In addition to simulating the detector response, the GEANT simulation also included 
the effects of pion decay.
In those kinematic bins where the acceptance was less than 10\%,
agreement between the empirical and Monte-Carlo-simulated acceptances
was poor.  Thus, an acceptance cut was applied such that only
kinematic bins that had acceptances greater than 10\%, and had no
neighboring bins with acceptances less than 10\%, were kept. In
addition to this ``10\% criterion,'' the bins at $\cosThetaCmPi > 0.9$
and $\cosThetaCmPi < -0.9$ were removed, since some portion of these
bins would have had acceptances of zero due to the geometry of
CLAS. The fraction of all kinematic bins rejected by the ``10\% criterion'' was
0.195.

The empirically-measured and Monte-Carlo-simulated acceptances agreed
well when these conditions were applied.  To quantify this agreement,
an ``acceptance ratio'' was determined, defined as the ratio of the
empirical acceptance to the Monte Carlo simulated acceptance for each
photon energy and $\cosThetaCmPi$ bin (with the acceptance cut
applied).  These acceptance ratios were placed in a histogram, and
then fit with a Gaussian.  The center of the Gaussian was 0.9997 and
the standard deviation was 0.040, which affirms the validity of the
Monte Carlo simulation of the response of CLAS to $\pi^+$.

In addition to examining the ratio of the empirical acceptance to the
Monte Carlo acceptance, 
a standardized Gaussian distribution $z_{ij}$ was created by
forming, for each kinematic bin, the difference of the 
Monte Carlo simulated acceptance $\epsilon_{MC}$ and the
empirical acceptance $\epsilon_E$, with that difference divided
by the combined acceptance uncertainty thusly:
\begin{equation}
z_{ij} = \frac{\left( \epsilon_{ij} \right)_{MC} - \left( \epsilon_{ij} \right)_{E}}
{\left( \sigma_{ij}  \right)_{MC+E}},
\label{eqn:z1}
\end{equation}
where
\begin{equation}
\left( \sigma_{ij}  \right)_{MC+E} =
\sqrt{\left( \sigma_{ij}  \right)^2_{MC} + \left( \sigma_{ij}  \right)^2_{E}}
\end{equation}
histogramed for each energy $i$ and $\cosThetaCmP$ $j$ kinematic 
bin.  These points are assumed to obey Gaussian statistics with a
variance of one and a centroid located at exactly zero.  If the
centroid of the distribution has been ``pulled'' away from zero, that
suggests the Monte Carlo acceptance results
$\left(\epsilon_{ij}\right)_{MC}$ do not approximate the empirical
acceptance exactly.  If the variance of the $z_{ij}$ distribution is less
than one, then the uncertainties $\left( \sigma_{ij}
\right)_{MC+E}$ are too large. Conversely, if the variance of
the $z_{ij}$ distribution is greater than one, this suggests the
uncertainties $\left( \sigma_{ij} \right)_{MC+E}$ are
underestimated.

The uncertainties of the Monte Carlo acceptance are assumed to be 
well-represented by the uncertainty appropriate for a binomial distribution:
\begin{equation}
\left( \sigma_{ij} \right)_{MC} =
\sqrt{\frac{\left( \epsilon_{ij} \right)_{MC} 
\left( 1 - \left( \epsilon_{ij} \right)_{MC}  \right)}
{\left( N_{ij} \right)_{Thrown}} },
\end{equation}
where $\left( N_{ij} \right)_{Thrown}$ is the number of events thrown in the 
$ij$ kinematic bin.

The mean of the standardized Gaussian
distribution $z_{ij}$
was from equation (\ref{eqn:z1}) nearly equal to zero within uncertainties, $0.10 \pm 0.09$.  
The value
of $\chi_{reduced}^2$ for the Gaussian fit to the distribution,
$\chi^2_{reduced} = 0.86$, was also reasonable.
However, the standard deviation, $1.29 \pm 0.09$, was larger than the
optimal value of one, suggesting that the uncertainties 
$\sigma_{MC+E}$
were too small.
When an additional 2\% uncertainty was added in
quadrature to the Monte Carlo uncertainty, the centroid, standard deviation, and $\chi_{reduced}^2$
were $0.09 \pm 0.07$, $1.02 \pm 0.05$, and $0.514$, respectively. 

To test
how far the Monte Carlo results were from optimal, we added 0.1\% to 
the Monte Carlo efficiency. With this 0.1\% shift to the Monte Carlo
the centroid, standard deviation, and $\chi_{reduced}^2$
were ($0.05 \pm 0.06$, $1.02 \pm 0.05$, $0.465$, respectively). 
Since the values are consistent with the optimal values, we assume 
henceforth that
the Monte Carlo acceptances agree very well with the empirical
acceptances when 2\% additional uncertainty is added to the Monte
Carlo acceptances.
The remainder of this analysis assumes that it is appropriate to
add this extra 2.0\% uncertainty in quadrature to the Monte Carlo
uncertainties on a bin-by-bin basis, and that has been done 
for each kinematic bin.

Having confirmed the validity of the Monte Carlo representation of the
CLAS response to $\pi^+$, the acceptance results for the reaction
$\piplusRxn$ were obtained by generating $10^7$ events (weighted
by the cross sections given by the SAID solution \cite{dugger}) that
were then processed in the same manner as the $\ppipiRxn$ comparison
reaction.  These simulated acceptances were used to determine the
differential cross sections reported here.

\subsection{Sector-by-sector comparison}

A sector-by-sector comparison of the differential cross sections was
performed to check the consistency of the extracted cross sections.
CLAS is constructed from six sectors which, ideally, should be
identical.  However, operationally, the response of each sector is
different owing to various hardware circumstances, problems, and
differences.  The simulations described in the previous section
incorporate knowledge of the various differences in the sectors in
order to properly reproduce the CLAS response for each particle type.
Since the Monte Carlo simulations should reflect sector-by-sector
changes in the detector arising from, for example, holes in the drift
chamber system due to broken wires and bad time-of-flight paddles, a
sector-by-sector comparison of the differential cross sections
inferred from the data obtained explores the reliability of the Monte
Carlo with respect to these detector irregularities.  The results of
this comparison indicated that variations attributable to
sector-by-sector variations were less than 0.4 \%, and much 
smaller than the uncertainty in the cross sections,
thus confirming the validity of the simulated 
sector-by-sector response.

A standardized Gaussian distribution for the sector-by-sector
comparison was created by forming, for each photon energy, $\cosThetaCmPi$,
and sector bin, the difference of the differential cross section in
each sector to the sector average and dividing the result by the
uncertainty.

The resulting centroid, standard deviation, and $\chi^2_{reduced}$ 
of the standardized Gaussian distribution were 
0.047 $\pm$ 0.021,  0.979 $\pm$ 0.018, and 1.01, respectively.
Thus, while the $\chi^2_{reduced}$
and standard deviation of the Gaussian are reasonable, the centroid
is somewhat smaller than the optimal value of zero.

To roughly estimate how far off the cross sections might be from the
desired value for the centroid, we shifted the sector average by a factor of
0.996. The resulting modified centroid, standard deviation, and $\chi^2_{reduced}$,
were found to be, 0.003 $\pm$ 0.021, 0.985 $\pm$ 0.018, and 0.979, respectively. 
Since this small shift of 0.4 \% in the sector average,
a shift much smaller than the uncertainty for the cross section,
produces parameters for the standardized Gaussian that are within optimal
values, the non-shifted parameters are acceptably close to
optimal.

\subsection{Bin migration}

To estimate the systematic uncertainty associated with bin migration,
the acceptance and efficiency results calculated using SAID-weighted
events were compared to acceptance and efficiency results using
non-weighted events.  Since the amount of the correction was found to
be typically less than 2\%, 
the systematic uncertainty associated
with bin migration was assumed to be ignorable.

\subsection{\label{sec-trigger}Trigger inefficiency}

The determination of a charged particle trigger inefficiency for the
{\tt{g1c}} data was performed by looking at data from a running period
just preceding the {\tt{g1c}} period, the {\tt{g2a}} running period.
(The {\tt{g2a}} running period is more fully described in
Ref. \cite{g2a}).  This running period had, in addition to the charged
particle trigger, a photon trigger. The photon trigger required that 
there was a hit in any two sectors of the electro-calorimeters located downstream 
of CLAS in coincidence with a hit in the photon-tagger.
By looking at {\tt{g2a}} events
that had a photon trigger and no charged trigger, yet had a $\pi^+$ in
the event, the inefficiency of the charged particle trigger in CLAS
for $\pi^+$ was determined.  This correction was applied to each
kinematic bin, and was always less than 1.0\%.
\subsection{Normalization}
\label{sec:Norm}


The {\em{absolute photon flux}} for the entire tagger photon energy
range was determined by measuring the rate of scattered electrons
detected in each counter of the focal plane of the bremsstrahlung
photon tagger by sampling focal plane hits not in coincidence with
CLAS.  The detection rate for the post-bremsstrahlung 
scattered electrons was integrated
over the lifetime of the experiment and converted to the 
corresponding
total number
of photons on target for each counter of the tagger focal plane.  The
tagging efficiency was measured in dedicated runs with a total
absorption counter (TAC) downstream of the cryogenic target, which
directly counted all photons in the beam. The details of the method
can be found in Ref. \cite{gflux}.

\section{Uncertainties}
\label{sec:Errs}

We summarize here the various uncertainties present in the cross
sections obtained in this work.

\begin{itemize}

\item An overall estimated systematic uncertainty of 1\% is taken as a
very conservative estimate of all sources of {\em{trigger inefficiency}}, 
as described in section~\ref{sec-trigger}.

\item The uncertainties associated with the {\em{detector response,
bin migration, and track reconstruction}} are contained within the
uncertainties associated with the Monte Carlo acceptance estimates
described in subsection~\ref{sec:Acceptance}.  These uncertainties are
taken into account on a bin-by-bin basis.

\item The uncertainties associated with the {\em{background
subtraction}} described in subsection~\ref{sec:Yield} are purely
statistical, and these were taken into account on a bin-by-bin basis.

\item The largest source of uncertainty in the {\em{photon flux
normalization}} arises from the uncertainty in the measurement of the
``tagger efficiency'' \cite{tag}, essentially a measure of the photon
beam collimation taken during normalization runs.  The value of this
tagger efficiency is dependent upon the positioning of the electron
beam supplied by the accelerator on the radiator of the photon tagger,
and will vary on a run-by-run basis determined by the run-by-run
condition of the electron beam tune.  With the procedure used to
obtain the photon flux normalization \cite{gflux}, the statistical
uncertainties associated with the photon flux normalization are always
far below 1\% and, when considered with other uncertainties in the absolute 
normalization, are negligable.

\item The systematic uncertainty of the {\em{absolute
normalization}} is comprised of six parts; three of them 
do not vary over the running period, while the remaining three do. 
The following quantities
vary over the running period:
\newcounter{bean5}
\begin{list}{\arabic{bean5}.}{\usecounter{bean5}\setlength{\rightmargin}{\leftmargin}}
\item run-to-run variations in the normalized neutron yield unaccounted
for by statistical uncertainties alone;
\item uncertainty in the target density \cite{targetDensity};
\item statistical uncertainty of the photon flux normalization.
\end{list}
Table \ref{tbl:systematic1} shows contributions to the systematic
uncertainties for quantities that varied over the running period. 

The following systematic uncertainties do not change over the running period: 
\newcounter{bean6}
\begin{list}{\arabic{bean6}.}{\usecounter{bean6}\setlength{\rightmargin}{\leftmargin}}
\item uncertainty in the liquid-hydrogen target-cell length, which was 0.3\% \cite{targetLength};
\item uncertainty associated with the tagger energy calibration
(described in subsection \ref{sec:Norm}), which was less than 1\%;
\item uncertainty in the trigger inefficiency correction, which was less than 1\%;
\end{list}
\end{itemize}

After adding all of the systematic uncertainties in quadrature,
the systematic uncertainty for
the absolute normalization is 1\%, 2\%, and 4\% for the 2.445
GeV, 3.115 GeV (full) and 3.115 GeV (half) data sets, respectively.
However, since combinations of more than one of these data sets was used
to obtain the differential cross section for each kinematic
bin, a 4\% absolute normalization uncertainty is assumed for simplicity.

\begin{table}[h]
\caption{Systematic uncertainties in the absolute normalization
for quantities that varied over the running period.
(The data set descriptions are discussed in section II.)
}\label{tbl:systematic1}
\vspace{2mm}
\centering
\begin{tabular}{|c|c|c|}
\hline
Data set &  Run-to-Run   &  Target density    \\
\hline
\multicolumn{1}{|c|}{2.445}
& 0.9\% & 0.1\%  \\
\hline
\multicolumn{1}{|c|}{3.115-full}
 & 1.9\% & 0.3\%  \\
\hline
\multicolumn{1}{|c|}{3.115-half}
& 3.6\% & 0.3\%  \\
\hline
\end{tabular}
\end{table}

\section{Results}
\label{sec:Results}

\begin{figure}[th]
\includegraphics[height=0.4\textwidth, angle=90]{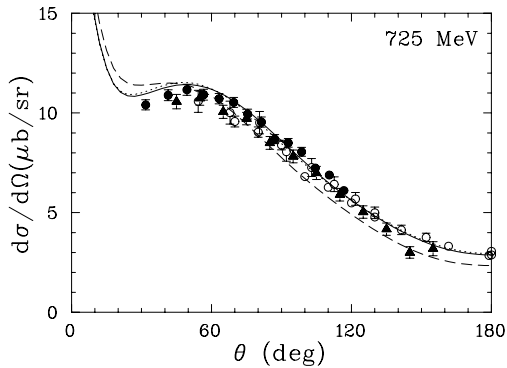}
\caption{The differential cross section for $\gamma
         p\to\pi^+n$ at E$_\gamma$ = 725~MeV versus pion center-of-mass
         scattering angle.  Solid (dotted) lines correspond
         to the SAID FA08 (FA07) solution. Dashed lines give the
         MAID07 \protect\cite{Maid07} predictions.  Experimental data
         are from the current (filled circles) and recent MAMI-B
         measurements (triangles) at
         723~MeV~\protect\cite{w21}. Previous bremsstrahlung
         measurements (open circles) are from
         Refs.~\protect\cite{w6,w8,w9,w14,w19}.  The
         data have been selected from energy bins
         spanning at most 3~MeV. Plotted uncertainties are
        statistical.
 \label{fig:g1}}
\end{figure}

The 618 differential cross sections obtained in this experiment are
compared to the world
data set~\cite{w1,w2,w3,w4,w5,w6,w7,w8,w9,w10,w11,w12,w13,w14,w15,w16,w17,w18,w19,w20,w21}
in Figs.\ \ref{fig:g1} $-$ \ref{fig:g3}, along with a number of
representative fits described below.  The differential cross sections
reported here are the first tagged $\pi^+n$ measurements above
780~MeV~\cite{w21}.
The cross sections are available in electronic form in
Ref.~\cite{diffcs}.  The database entries include the differential
cross sections, as well as uncertainties (excluding the overall
absolute normalization uncertainty), for each incident photon energy
and $\cosThetaCmPi$ bin shown in this paper.

For a specific example of agreement with previous measurements, in
Fig.\ \ref{fig:g1} we compare differential cross sections obtained
here with those from the A2 collaboration of the MAMI-B group~\cite{w21}, 
at an energy common to both
experiments.  The CLAS data and the results from MAMI-B appear to agree 
well at this energy.

More generally, as can be seen in Figs.~\ref{fig:g1} $-$ \ref{fig:g3},
agreement with previous measurements is good overall. The largest
deviations generally occur at forward angles. Thus further measurements at
more forward angles would be useful. 
While agreement with previous measurements is generealy good, 
even so, the data here extend
measurements to higher energies with more
complete angular coverage than obtained in those previous measurements.

\section{Amplitude Analysis of Data}
\label{sec:fit}


We have included the new cross sections from this experiment in a
number of multipole analyses covering incident photon energies up to
2.7~GeV using the full SAID database in order to gauge the influence
of the present measurements, as well as their compatibility with
previous measurements.  A ``forced" fit, which included the present
dataset weighted by an arbitrary factor of 4, was compared to a standard fit.
(The standard fit with normal weighting is called henceforth FA08.)  
The results with two different weightings
were in good agreement, despite the CLAS data
having a larger weighting. This agreement is not surprising
concidering the agreement of these new data with previously
published measurements and that an
older fit (FA07) was able to give a reasonable prediction for the
previously published cross sections at all but the highest
energies~\cite{SAID}.

\begin{figure*}[th]
\includegraphics[height=0.8\textwidth, angle=90]{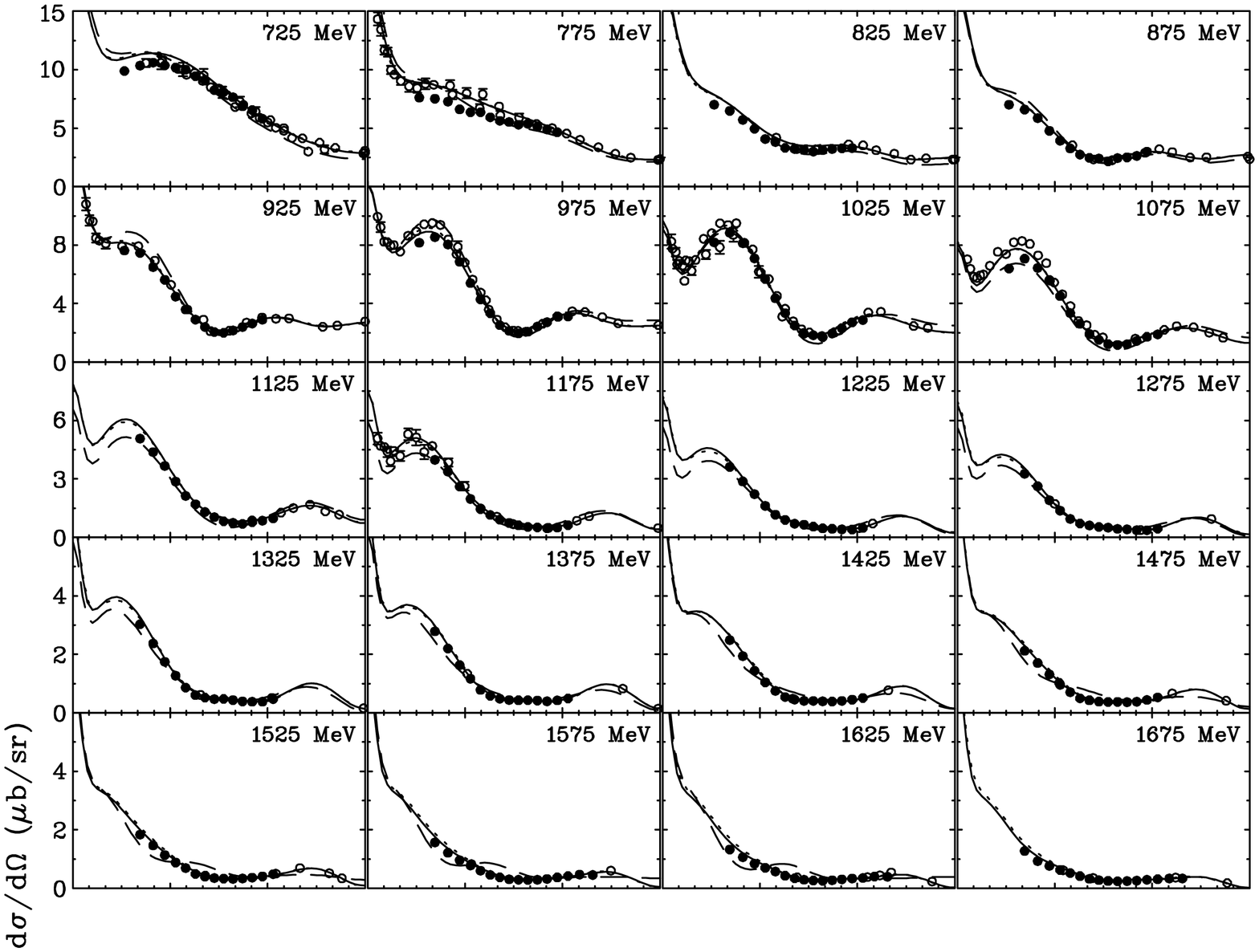}\\
\includegraphics[height=0.8\textwidth, angle=90]{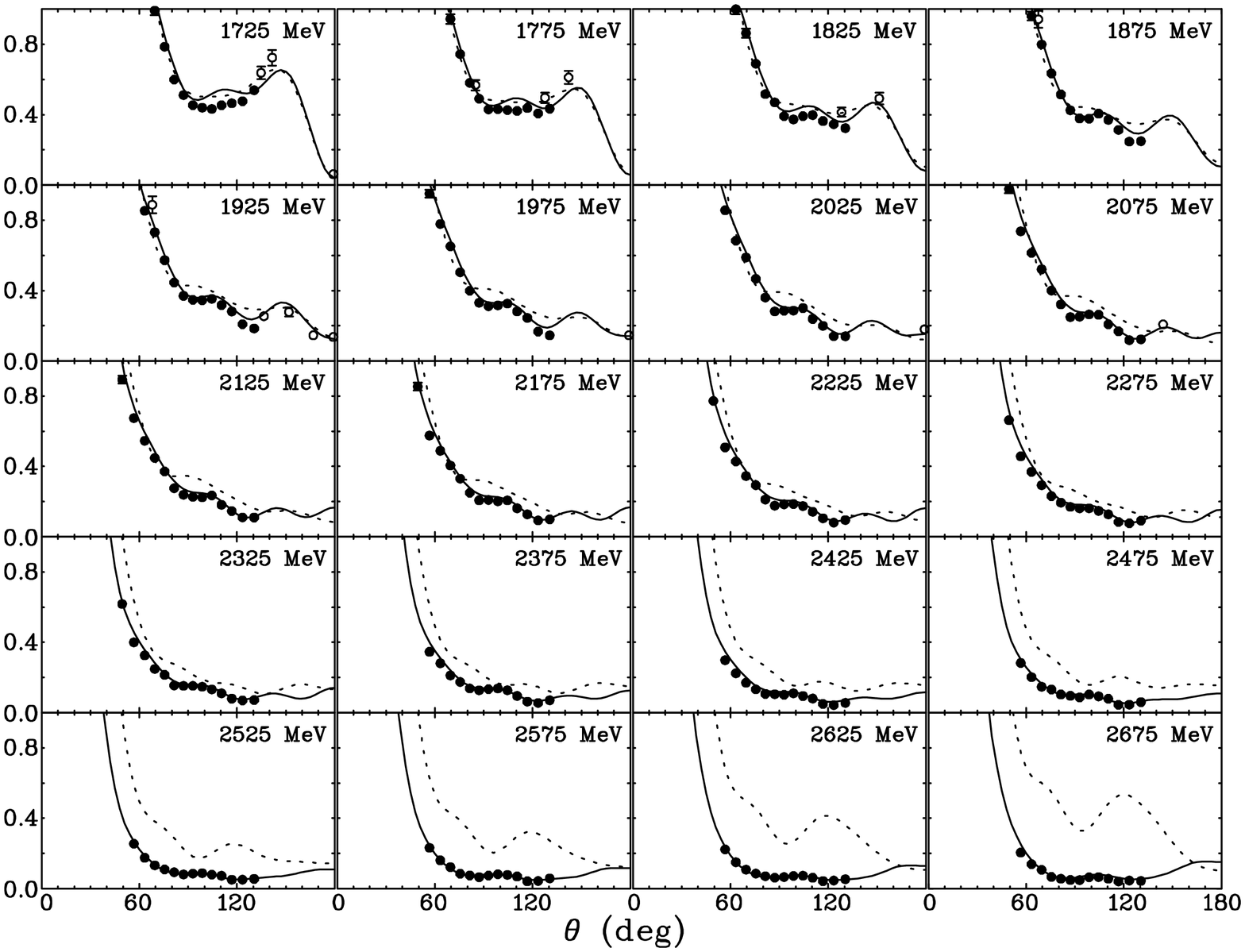}
\caption{The differential cross section for $\gamma 
         p\to\pi^+n$ below E$_\gamma$ = 2.7~GeV versus pion 
	center-of-mass scattering angle.  Solid (dotted) lines 
	correspond to the SAID FA08 (FA07) solution. Dashed lines 
	give the MAID07 \protect\cite{Maid07} predictions.  
	Experimental data are from the current (filled circles) 
	and previous measurements (open circles).  The plotted  
	points from previously published experimental data are 
	those data points within 3~MeV of the photon energy 
	indicated on each panel. Plotted uncertainties are
	statistical.
\label{fig:g2}}
\end{figure*}

\begin{figure*}[th]
\includegraphics[height=0.69\textwidth, angle=90]{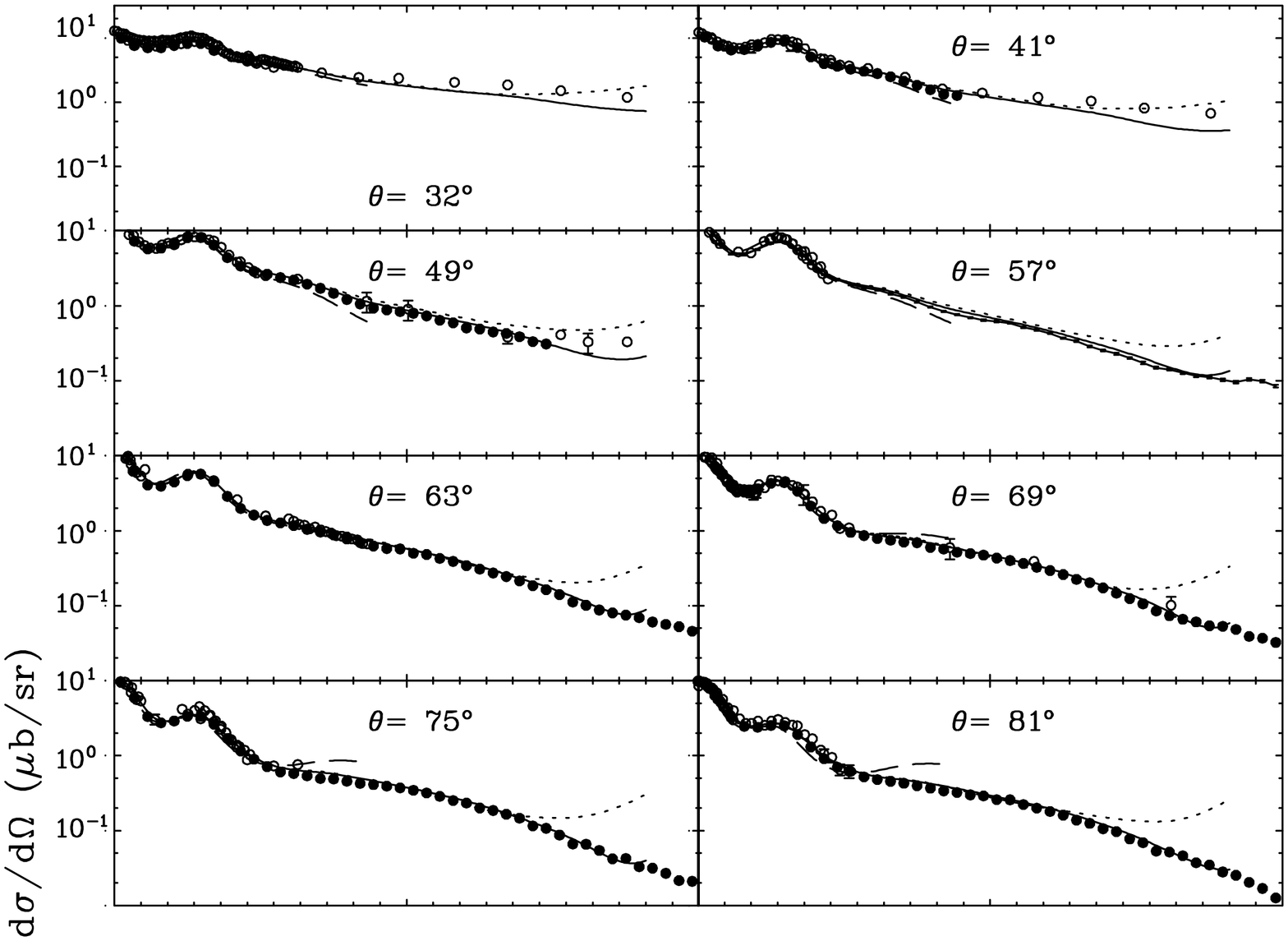}\\
\includegraphics[height=0.69\textwidth, angle=90]{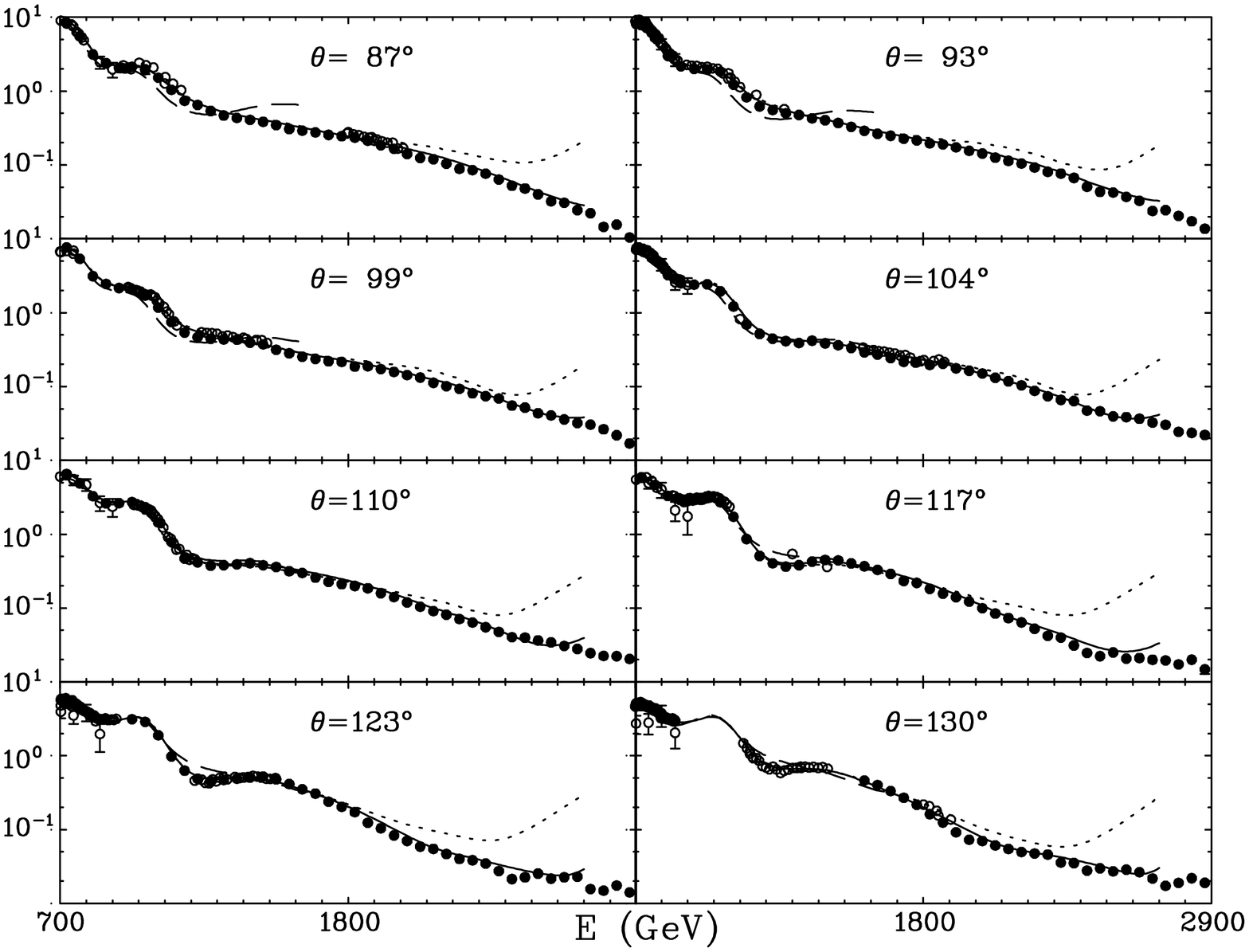}        
\caption{Fixed angle excitation functions for $\gamma               
        p\to\pi^+n$. The pion center-of-mass scattering angle   
        is shown. Notation as in Fig.  \protect\ref{fig:g2}.       
        The plotted  points from previously published
        experimental data are those data points within 2           
        degrees of the angle indicated on each panel.
        \label{fig:g3}}
\end{figure*}

In Table~\ref{tab:tbl1}, we compare FA08 with two previous SAID fits
(FA07 and FA06~\cite{dugger}) and also with the Mainz fit
MAID07~\cite{Maid07} up to its stated center-of-mass energy $W$ limit
of 2~GeV ($E_\gamma$ = 1.65~GeV).  The FA07 fit included LEPS
Collaboration $\pi^0p$ measurements~\cite{LEPS}.  These three
solutions are compared with the data in Figs.~\ref{fig:g2} and
\ref{fig:g3}. While the FA07 and FA08 SAID fits agree well over 
the energy range of the Mainz fit, disagreements between the SAID 
and MAID fits are most pronounced at angles more forward than the 
CLAS data. Near its upper energy limit, the
MAID07 solution also exhibits structure not seen in the data.

Above 2.4~GeV, the new CLAS data reported here begin to depart from
the FA07 predictions. As a result, the new data presented here have
resulted in adjustments of a number of parameters in the FA08 solution
so that the new solution better reproduces the measured cross sections, which are
significantly lower than the predictions given by FA07.

In fitting the data, the stated experimental systematic uncertainties
have been used as an overall normalization adjustment factor for the
angular distributions \cite{sm02}.  FA07 included all previously
published data used in FA06~\cite{dugger}, plus recent $\pi^0p$
differential cross sections and beam asymmetry $\Sigma$ data from the
LEPS Collaboration~\cite{LEPS}.  The MAID07 analysis does not include
the recent $\pi^0p$ measurements from CLAS \cite{dugger} and LEPS
\cite{LEPS}, and has a center-of-mass energy limit of $W$ = 2~GeV
($E_\gamma$ = 1.65~GeV).  Presently, the pion photoproduction database
below $E_\gamma$ = 2.7~GeV consists of 25639 data points that have been 
fit in the FA08 solution with $\chi^2$ = 54161.  The contribution to 
the total $\chi^2$ in the FA08 analysis of the 561 new CLAS $\pi^+n$ 
data points (e.g., those data points up to $E_\gamma =$ 2.7~GeV) is 
1407.

\begin{table}[th]
\caption{$\chi^2$ comparison of fits to pion photoproduction
	data up to 2.7~GeV.  Results are shown for three 
	different SAID solutions (FA08, FA07, and FA06)
	recent MAID07.  See text for details. Comparison 
	includes all previous plus new CLAS $\pi^+n$ 
	measurements. \label{tab:tbl1}}
\vspace{2mm}
\begin{tabular}{|c|c|c|c|}
\colrule
Solution & Energy limit & $\chi^2$/Data & Data \\
         & (MeV)        &               &      \\
\colrule
FA08     & 2700        &  2.11         & 25639 \\
FA07     & 2700        &  2.02         & 24376 \\
FA06     & 3000        &  2.15         & 25252 \\
MAID07   & 1650        &  7.38         & 22621 \\
\colrule
\end{tabular}
\end{table}

\begin{figure*}[th]
\centerline{
\includegraphics[height=0.45\textwidth, angle=90]{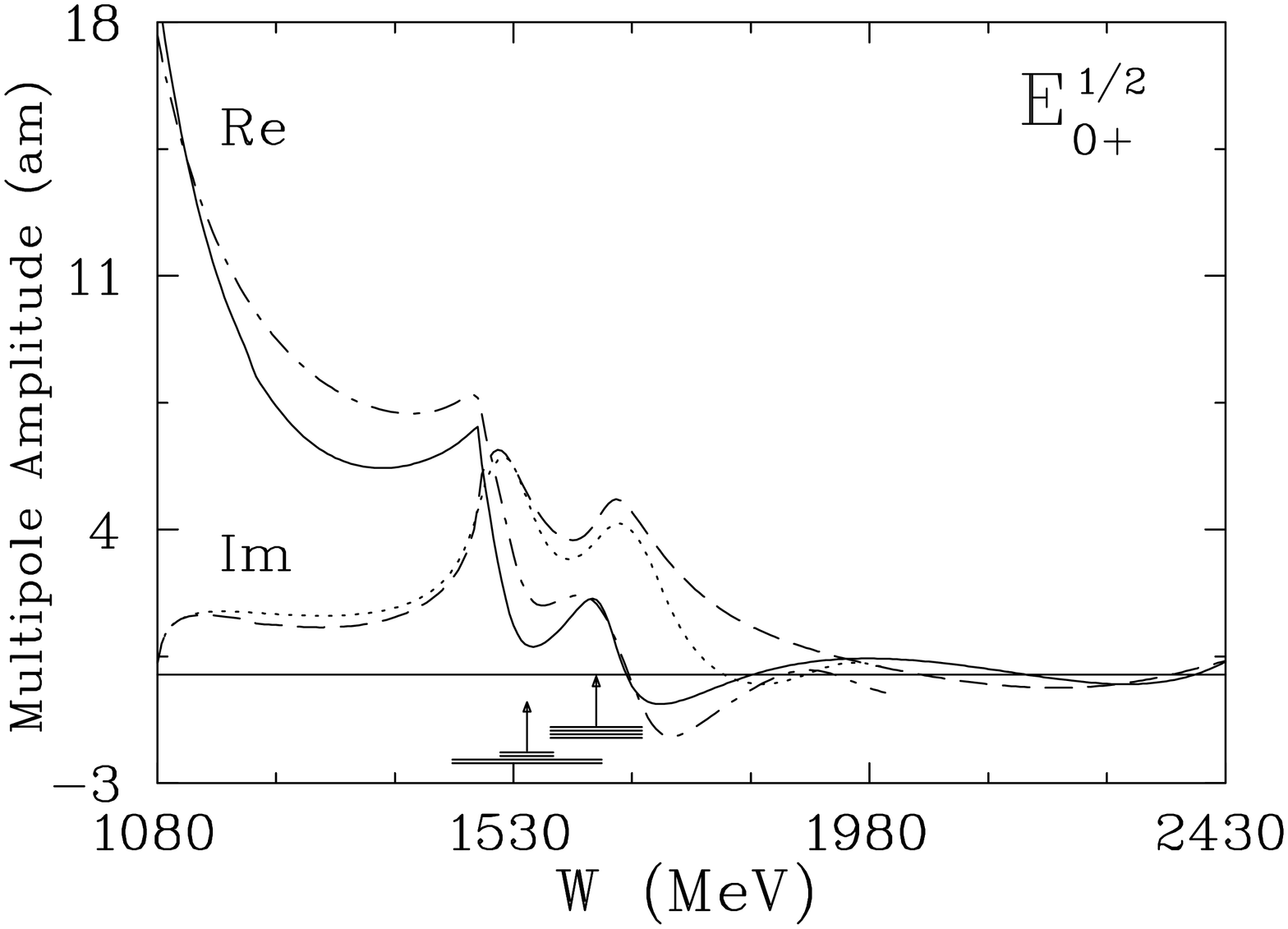}\hfill
\includegraphics[height=0.45\textwidth, angle=90]{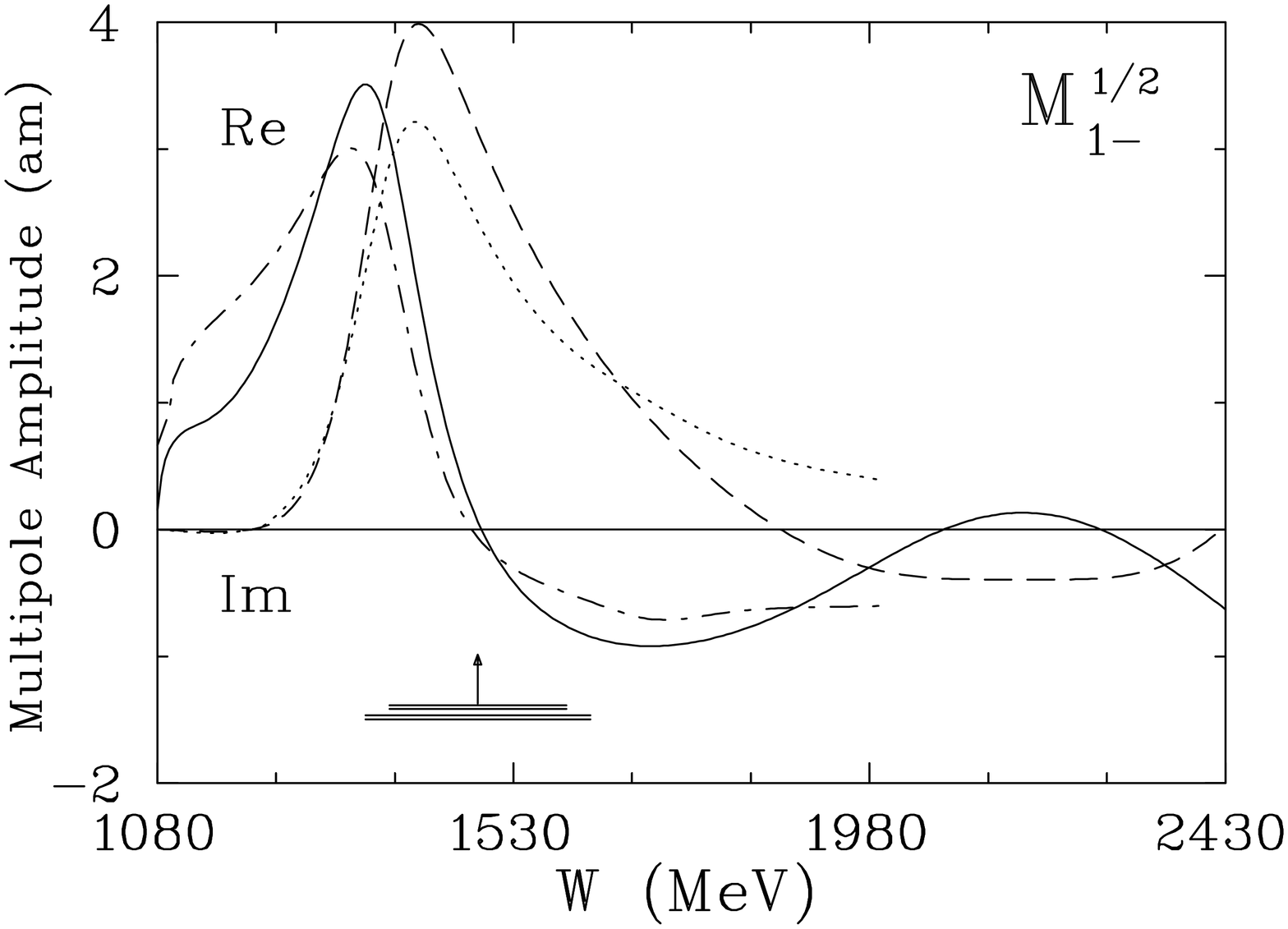}}
\centerline{
\includegraphics[height=0.45\textwidth, angle=90]{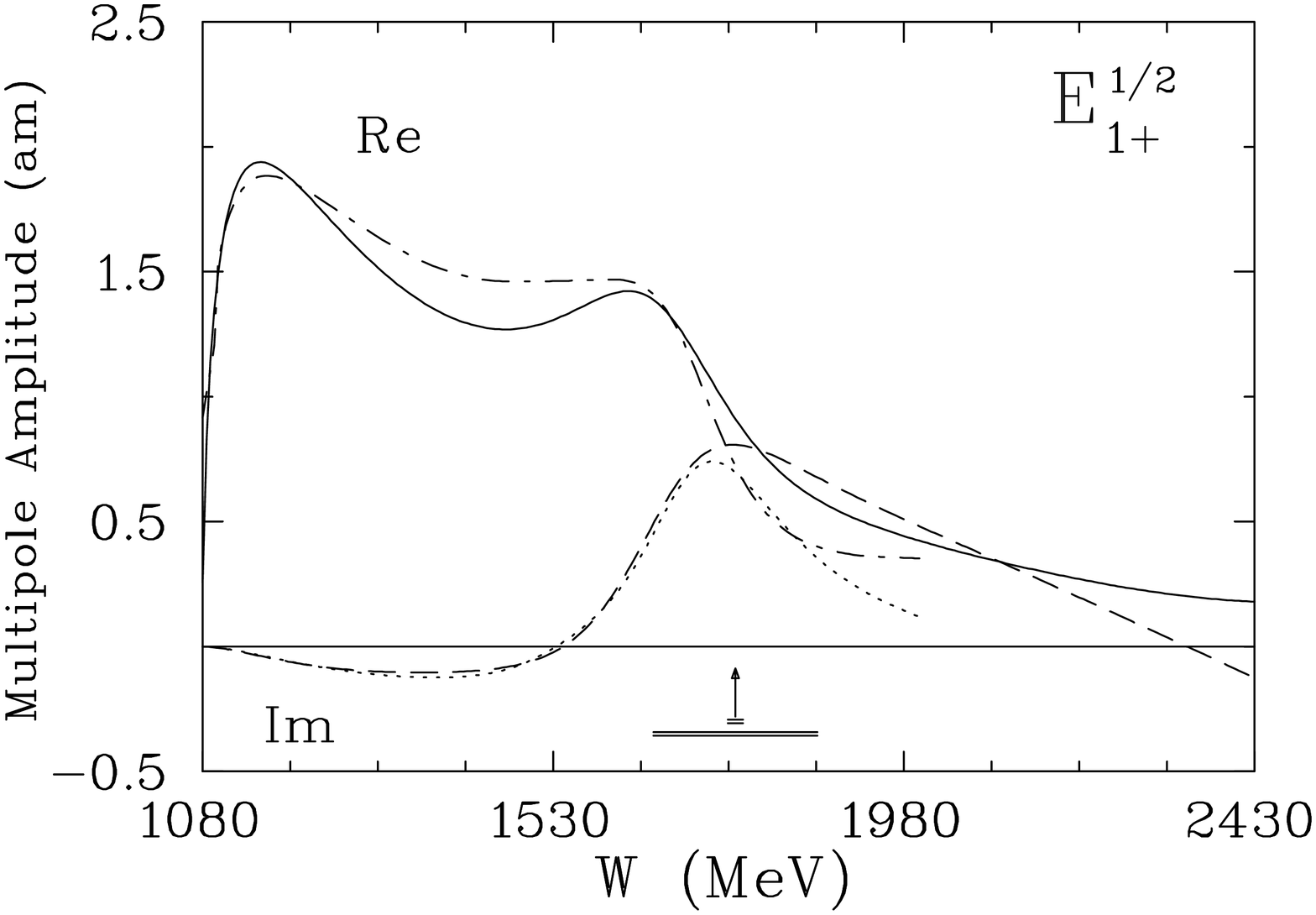}\hfill
\includegraphics[height=0.45\textwidth, angle=90]{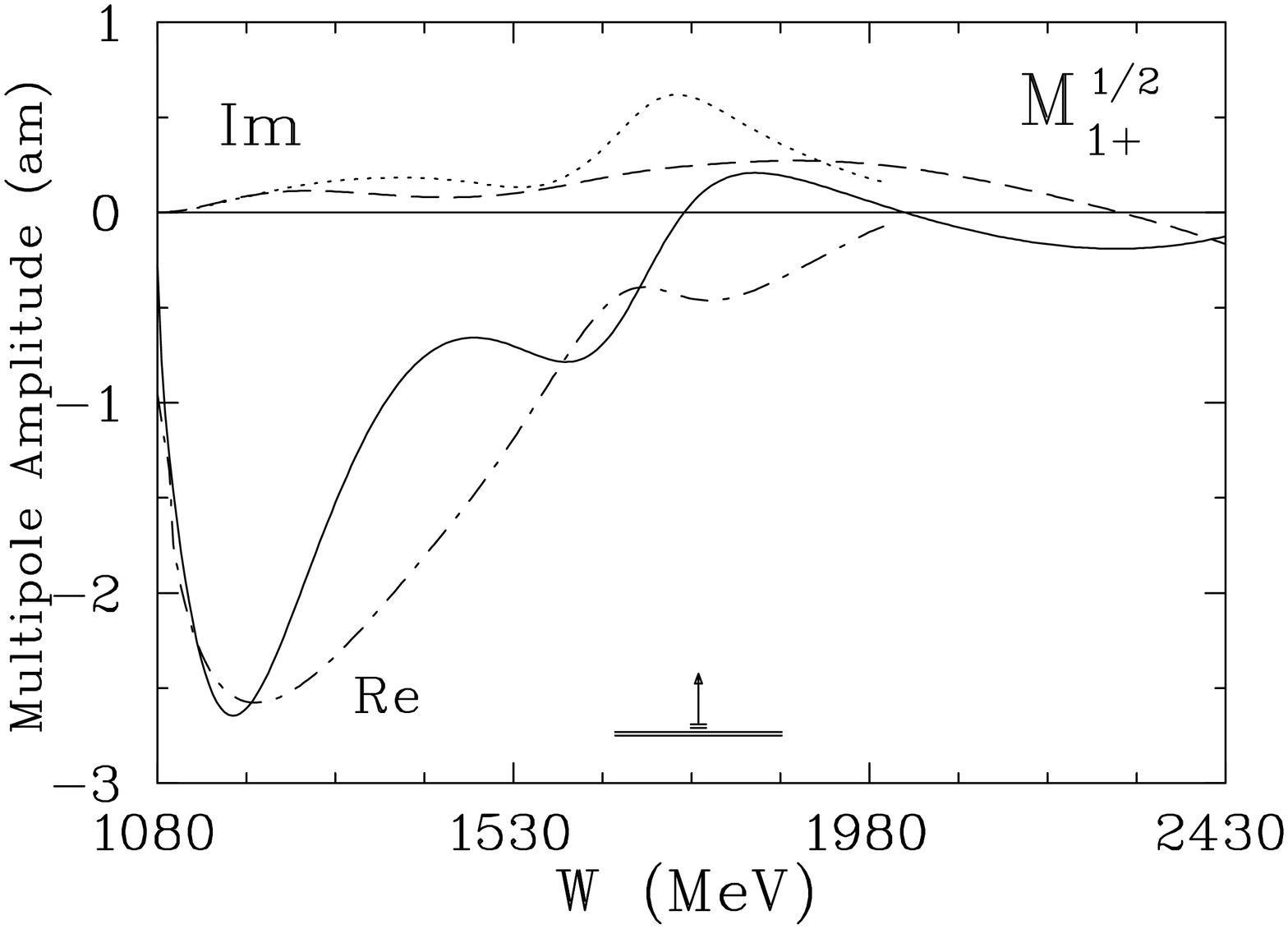}}
\centerline{
\includegraphics[height=0.45\textwidth, angle=90]{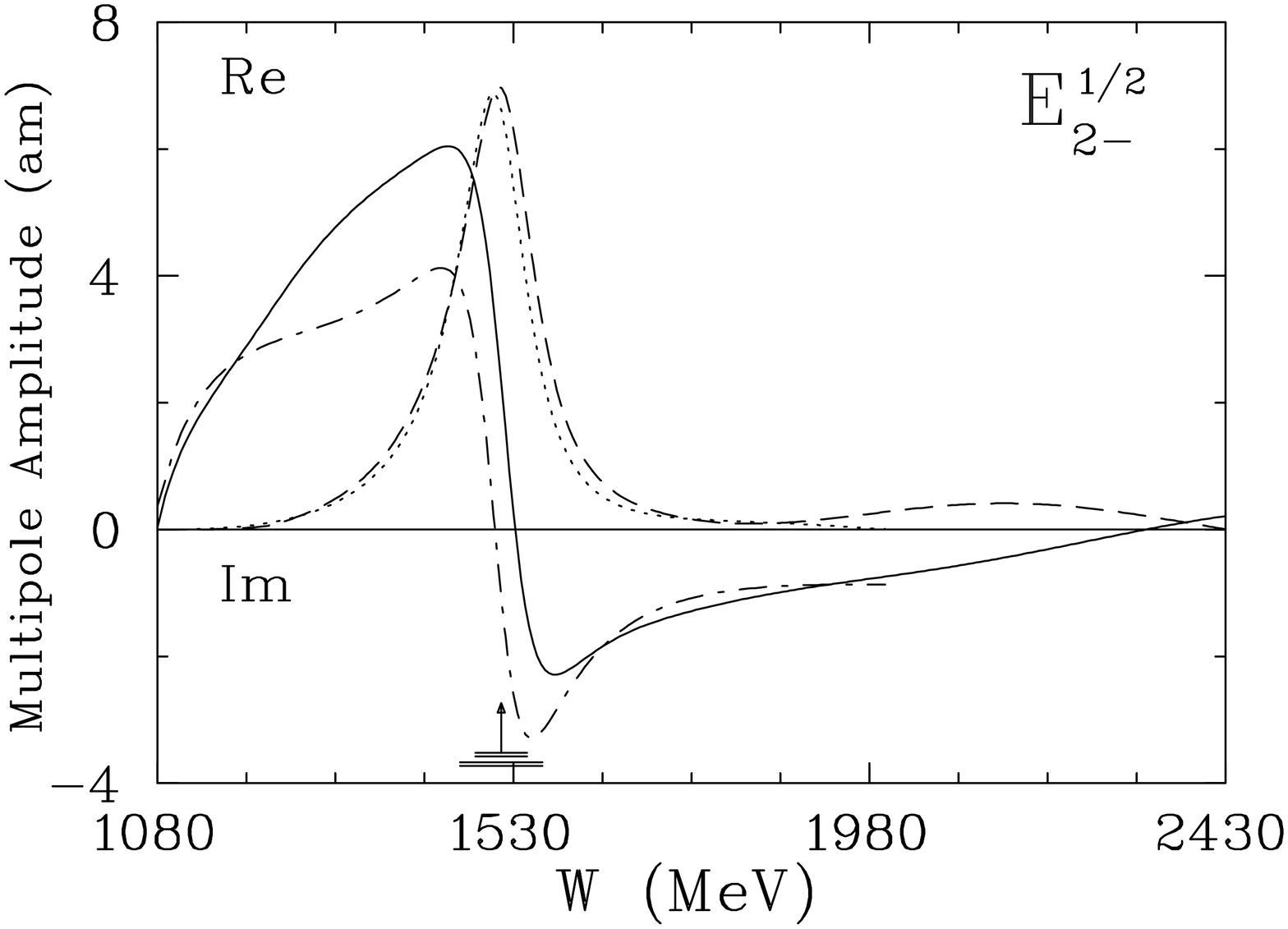}\hfill
\includegraphics[height=0.45\textwidth, angle=90]{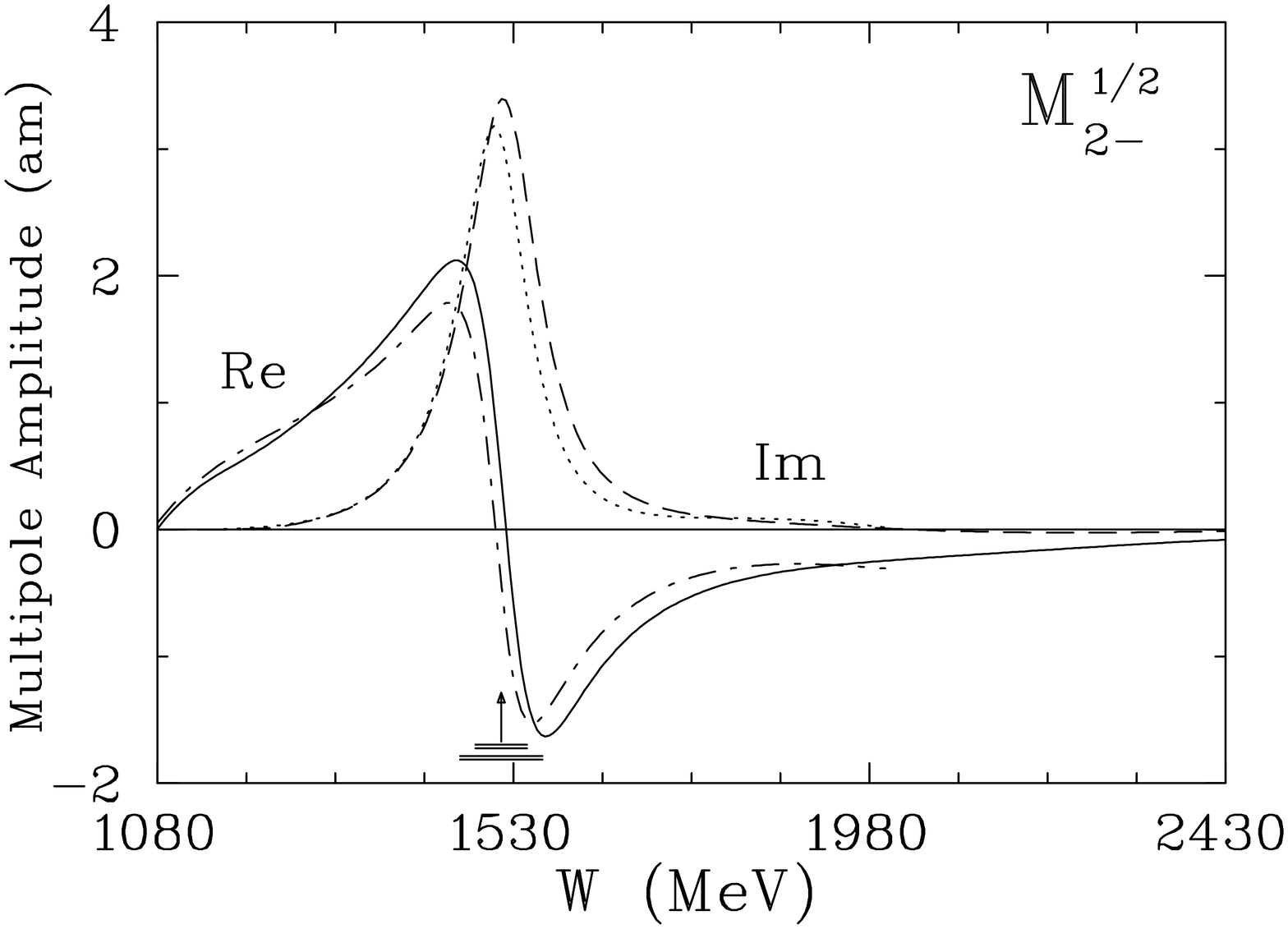}}
\centerline{
\includegraphics[height=0.45\textwidth, angle=90]{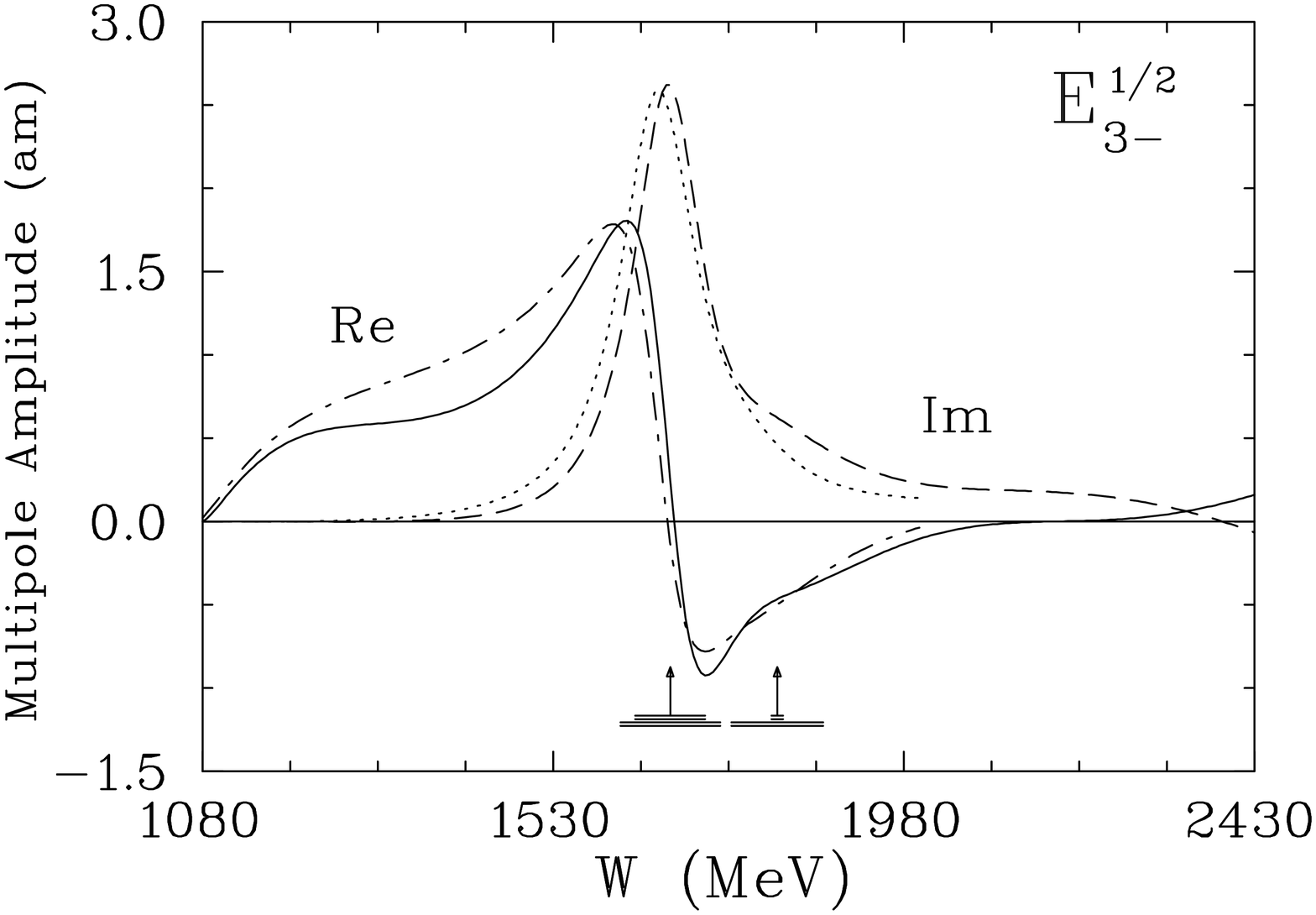}\hfill
\includegraphics[height=0.45\textwidth, angle=90]{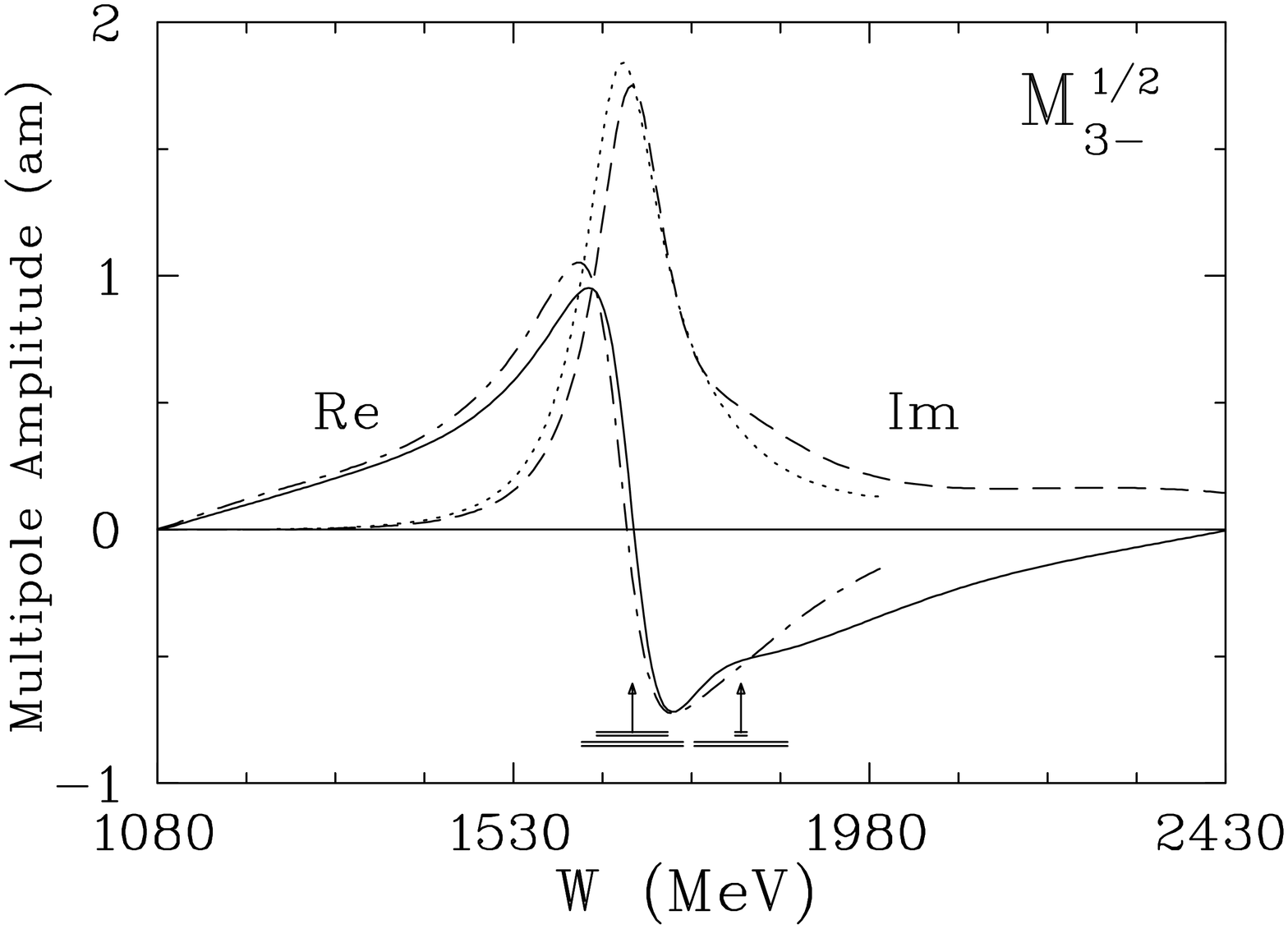}}
\caption{Multipole amplitudes from threshold to $E_{\gamma}$
         = 2.7~GeV for isospin $1/2$.  Solid (dashed) lines correspond
         to the real (imaginary) part of the FA08 solution.  Dashed-dot (dotted)
         lines give real (imaginary) part of the MAID07~\protect\cite{Maid07}
         solution.  Vertical arrows indicate $W_R$ and horizontal bars
         show full $\Gamma$ and partial widths for $\Gamma_{\pi N}$
         associated with the SAID $\pi N$ solution
         SP06~\protect\cite{sp06}.  \label{fig:g6}}
\end{figure*}

\begin{figure*}[tl]
\centerline{
\includegraphics[height=0.45\textwidth, angle=90]{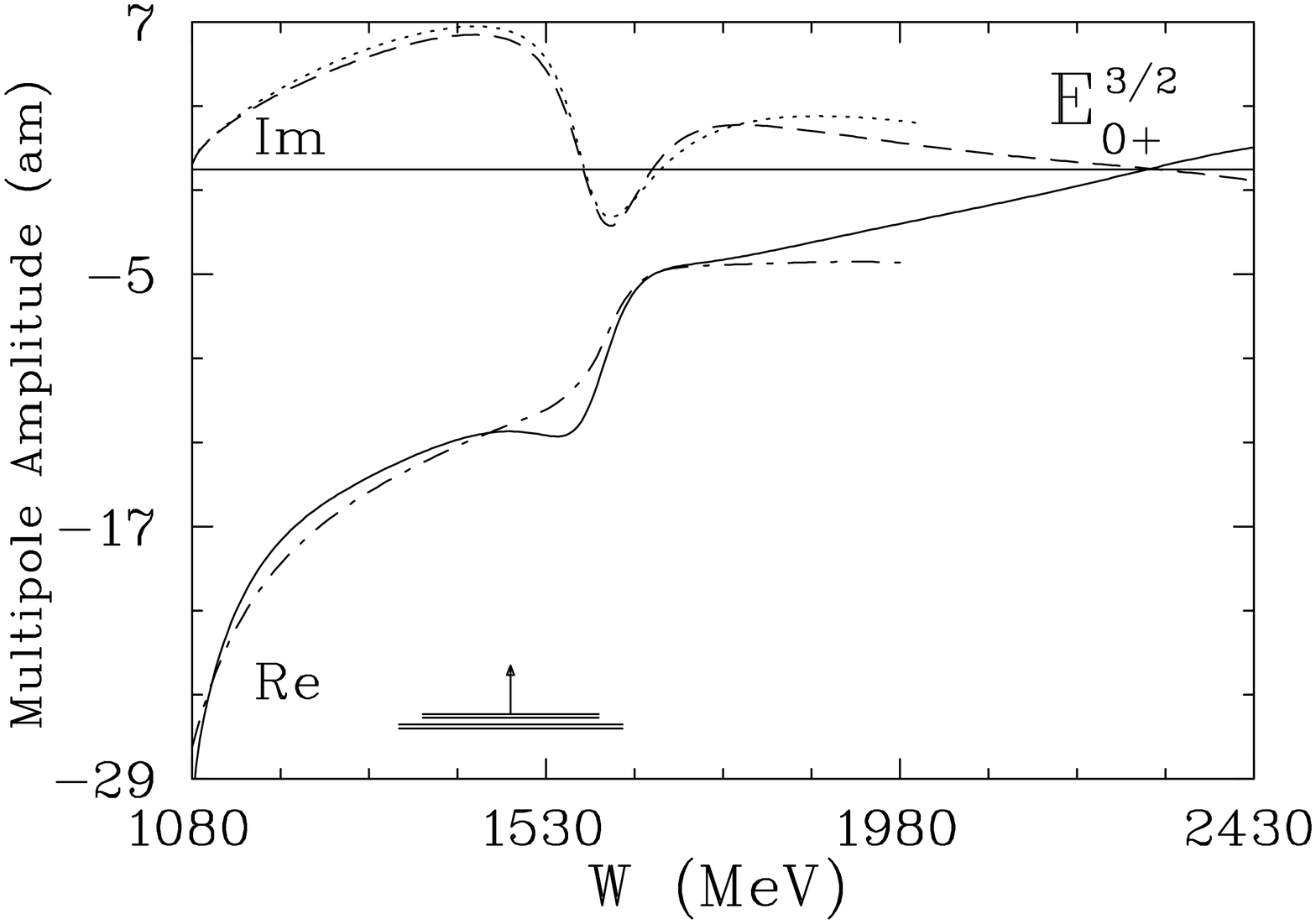}\hfill
\includegraphics[height=0.45\textwidth, angle=90]{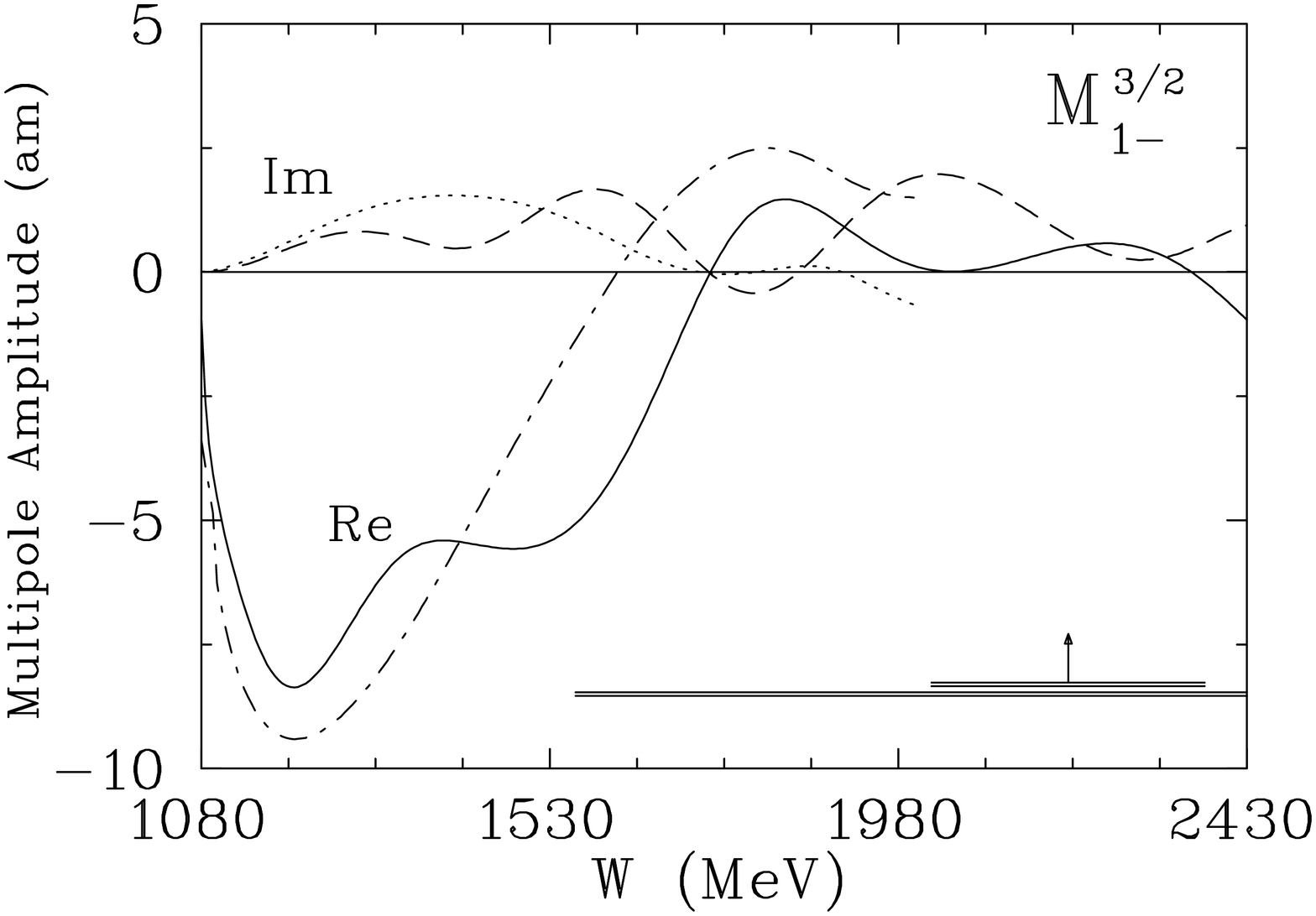}}
\centerline{
\includegraphics[height=0.45\textwidth, angle=90]{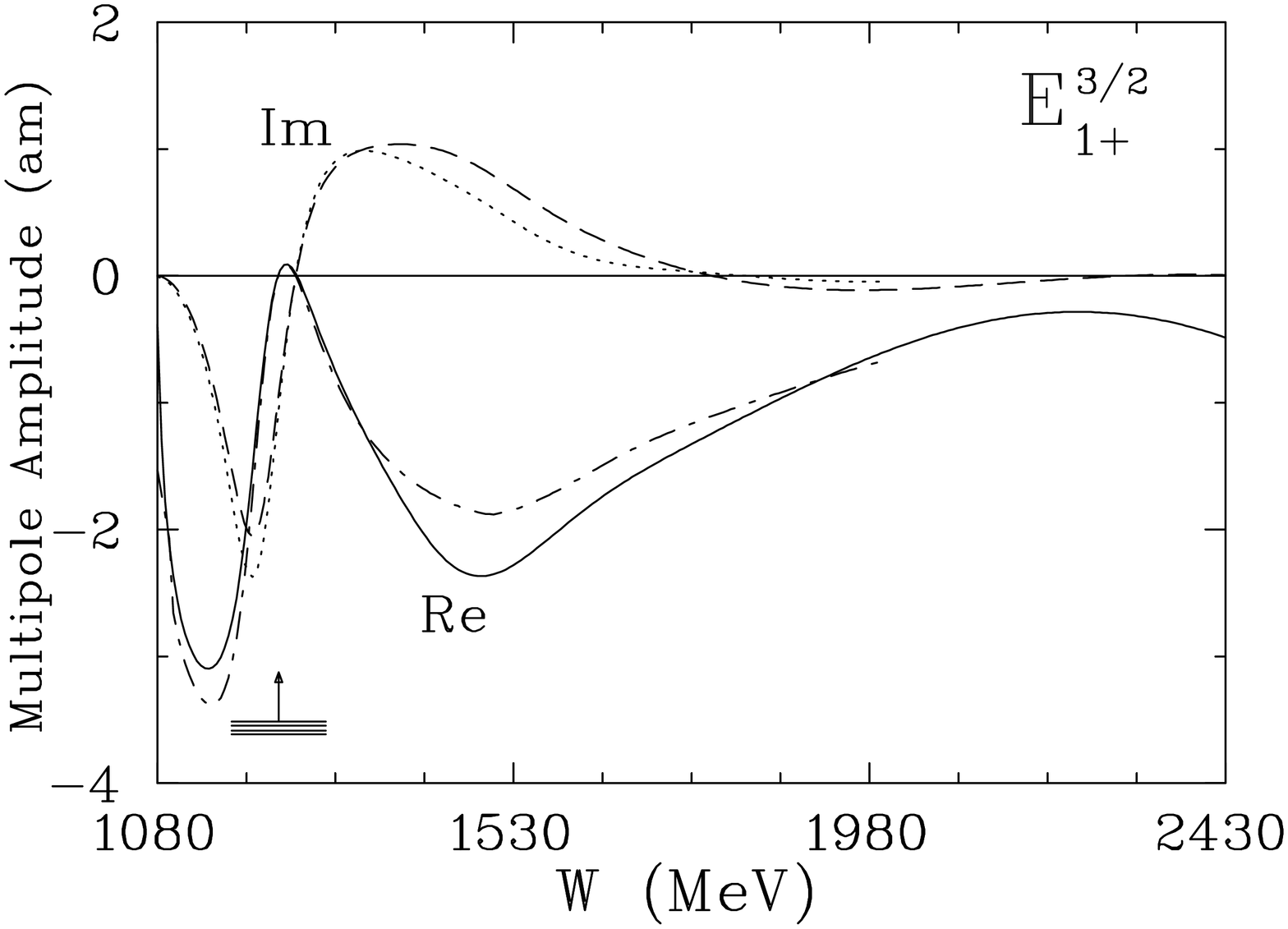}\hfill
\includegraphics[height=0.45\textwidth, angle=90]{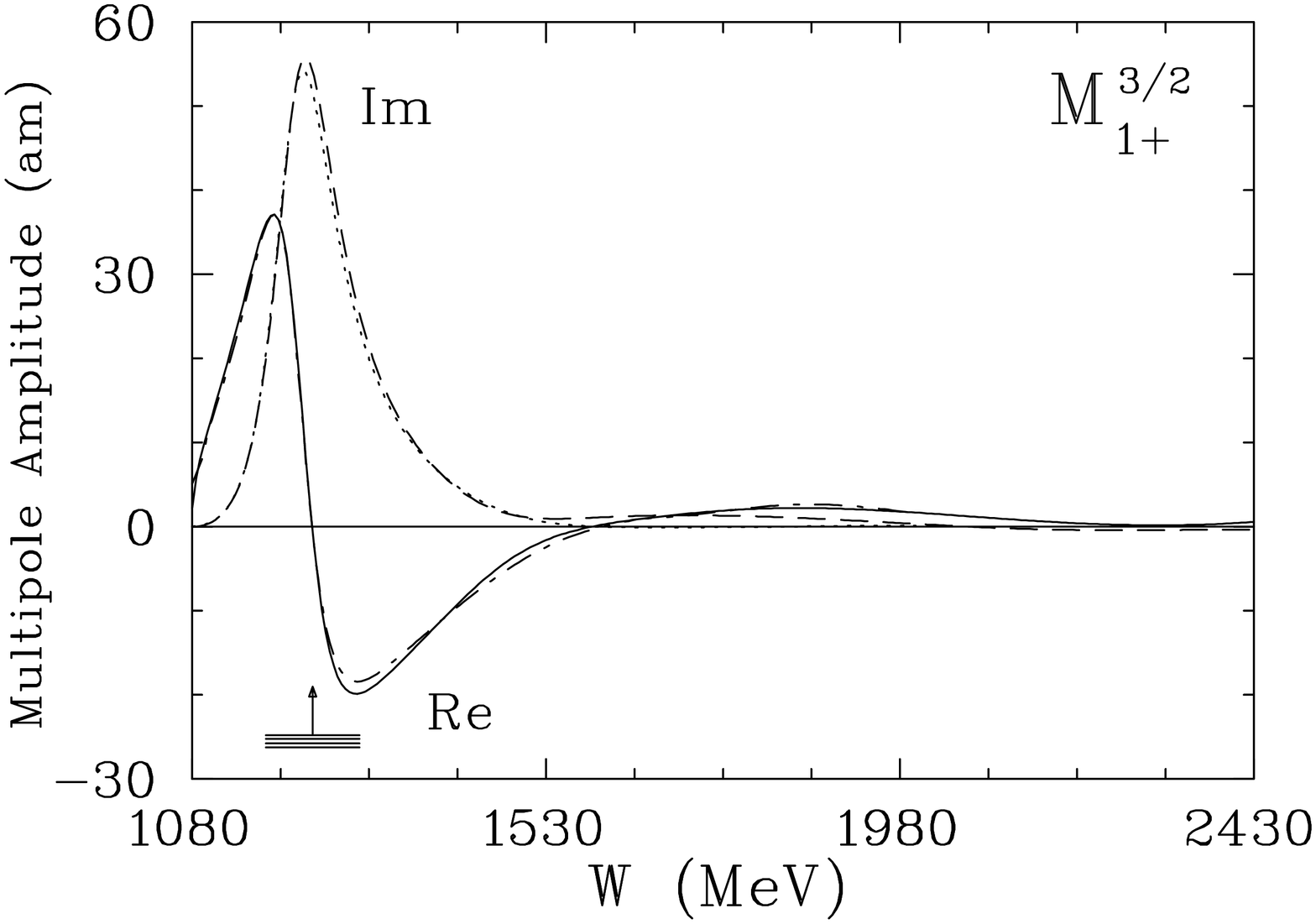}}
\centerline{
\includegraphics[height=0.45\textwidth, angle=90]{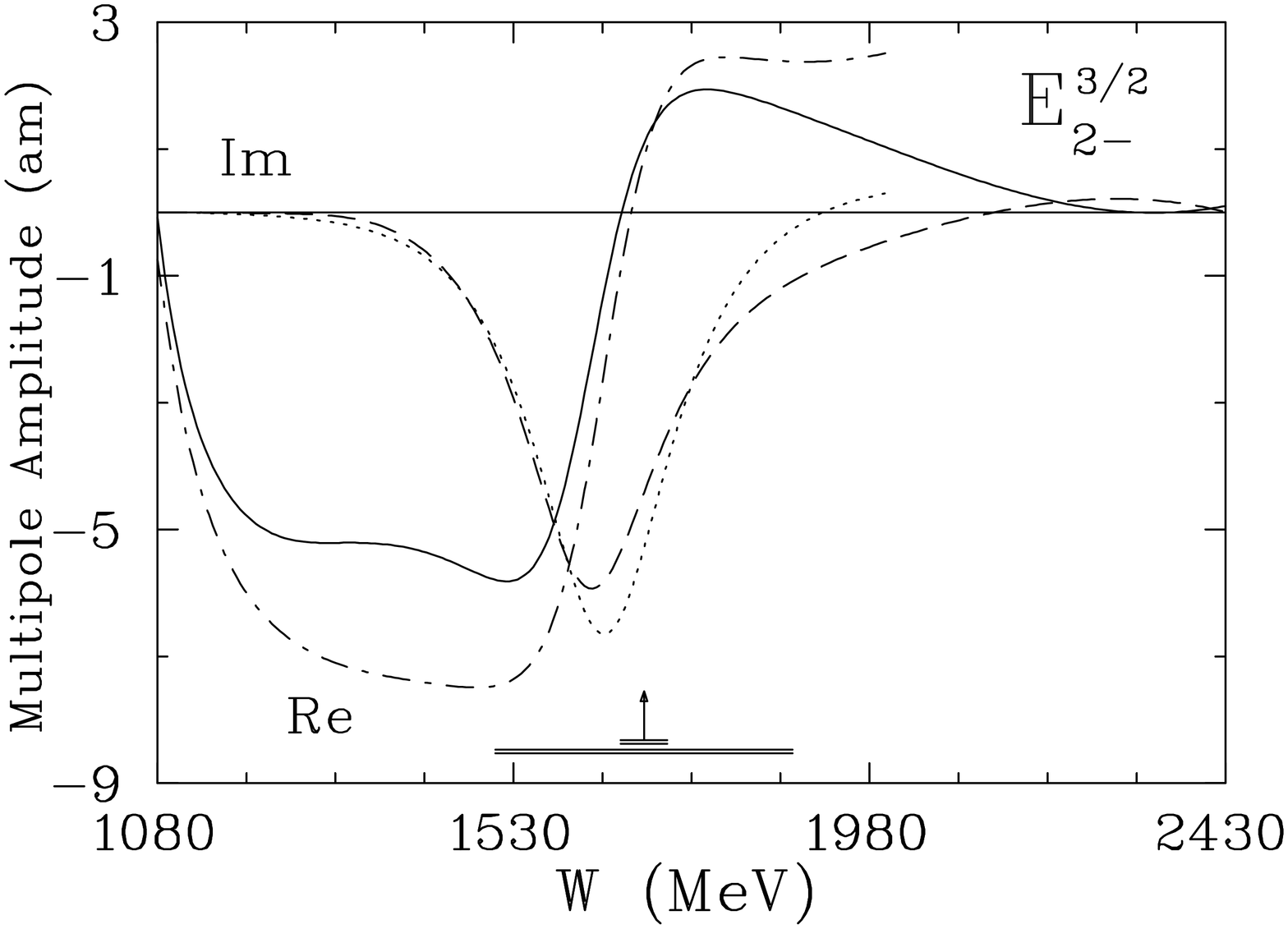}\hfill
\includegraphics[height=0.45\textwidth, angle=90]{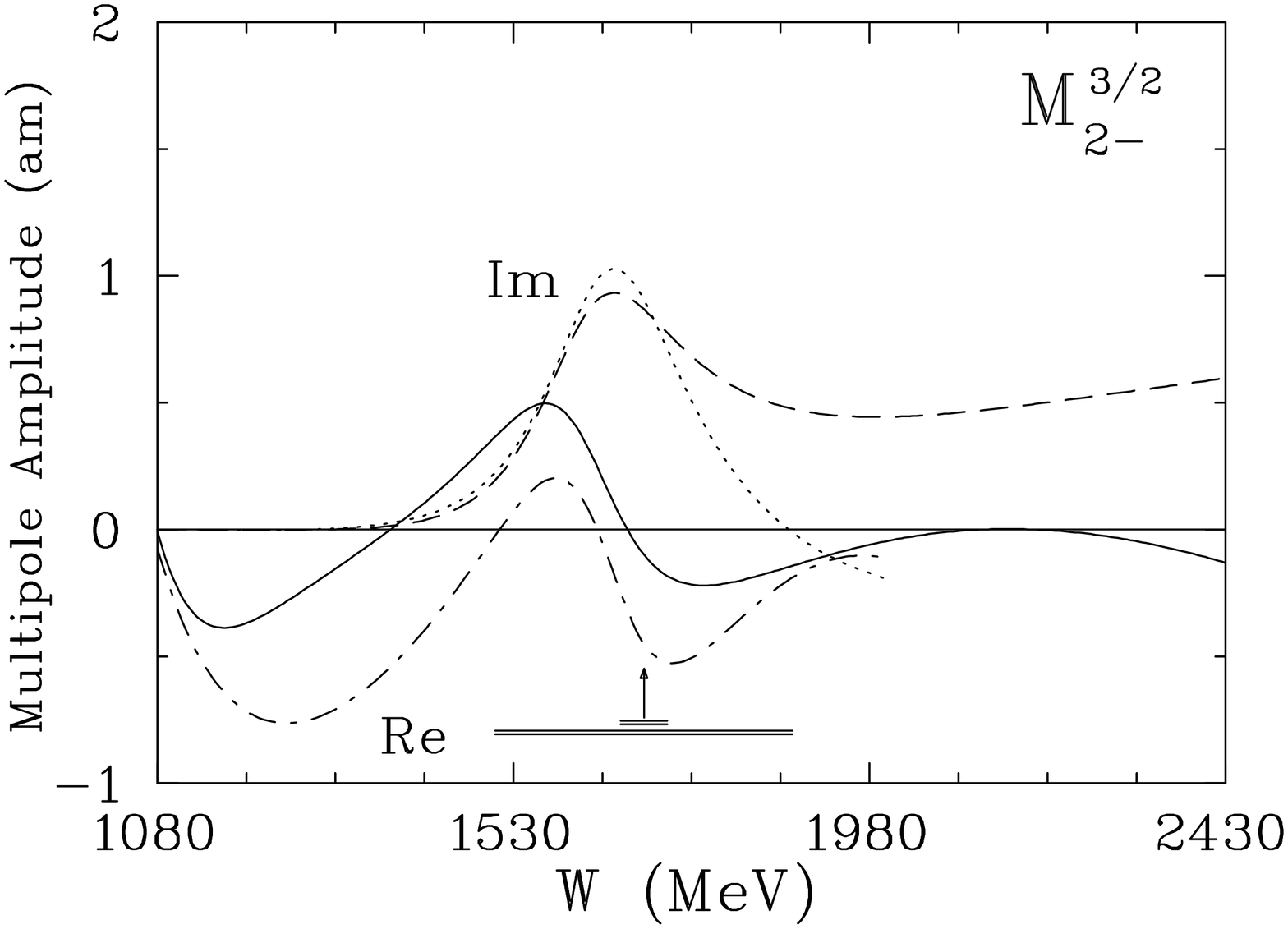}}
\centerline{
\includegraphics[height=0.45\textwidth, angle=90]{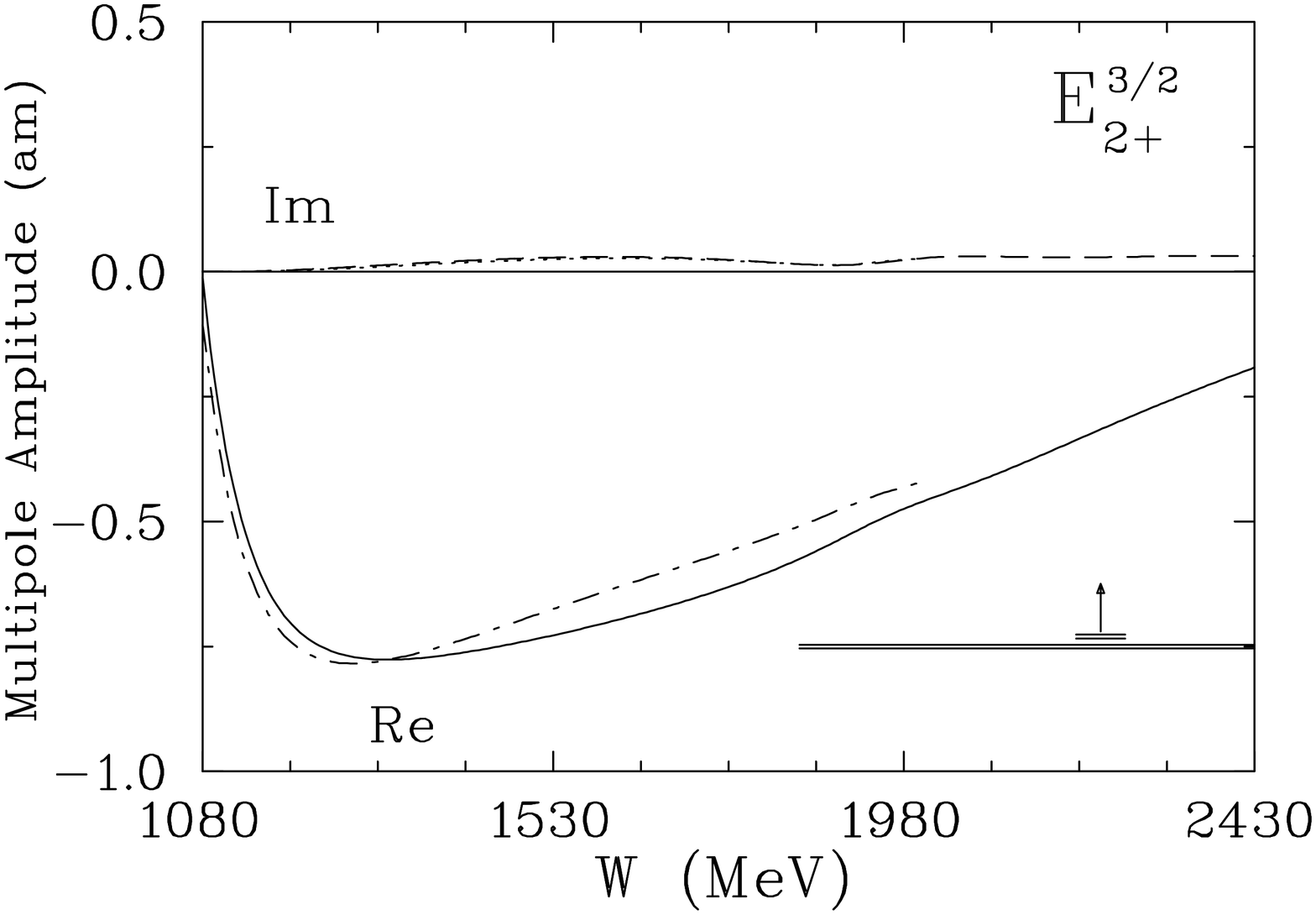}\hfill
\includegraphics[height=0.45\textwidth, angle=90]{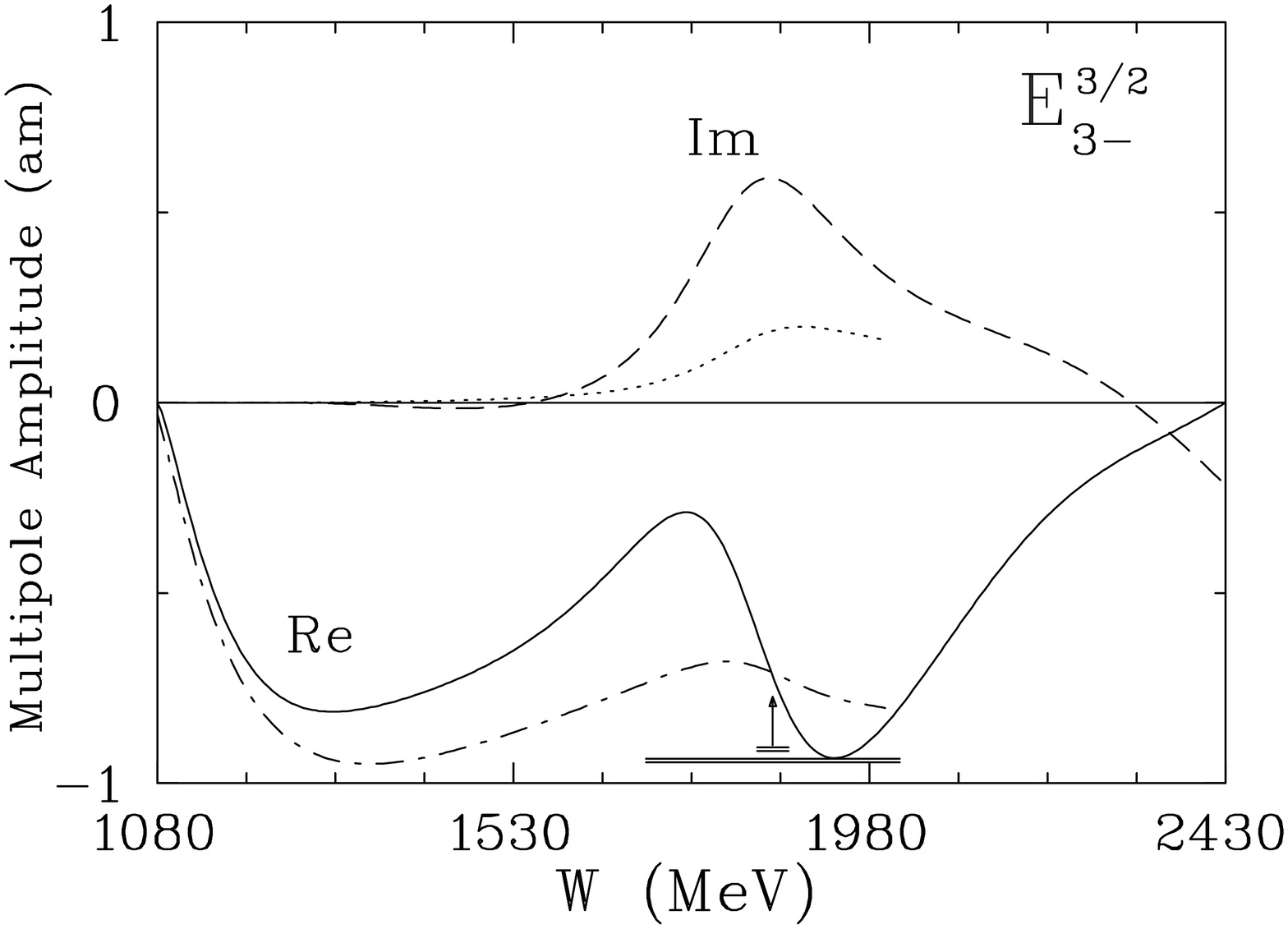}}
\caption{Multipole amplitudes from threshold to $E_{\gamma}$
         = 2.7~GeV for isospin $3/2$.  Notation as in
         Fig.~\protect\ref{fig:g6}. \label{fig:g7}}
\end{figure*}

Multipoles from the FA08 fit are compared to the earlier MAID07
determinations in Figs.~\ref{fig:g6} and \ref{fig:g7}.  Both FA07 and
FA08 are quite similar, but significant differences 
between SAID and MAID in magnitude (e.g., $E^{1/2}_{2-}$,$M^{3/2}_{2-}$,
and $E^{3/2}_{3-}$) and $W$ dependance (e.g., $M^{1/2}_{1+}$, and $M^{3/2}_{1-}$)
are seen. Given that large differences are
not seen in the differential cross sections, further measurements of spin observables
will be needed to better constrain these amplitudes.

\begin{figure}[th]
\includegraphics[height=0.4\textwidth, angle=90]{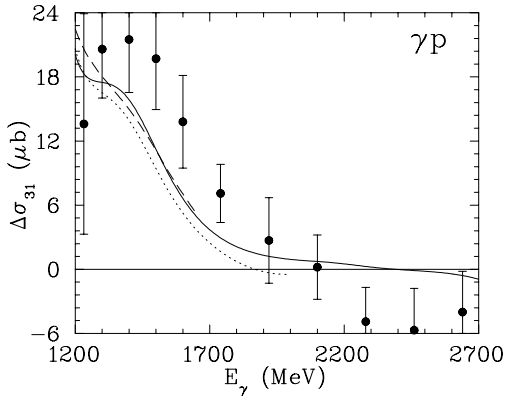}
\caption{Single-pion photoproduction contributions to the
          proton GDH sum rule $\Delta\sigma_{31} = \sigma_{3/2} -
          \sigma_{1/2}$ from the SAID current (solid), recently
          published SM05~\protect\cite{GDH} (dotted), and
          MAID07~\protect\cite{Maid07} (dashed) analyses. GDH data
          from Ref.~\protect\cite{Dutz}.  Plotted uncertainties are
          statistical and systematical added in quadratures.
\label{fig:g4}}
\end{figure}

\clearpage
With the addition of CLAS $\pi^0p$ and $\pi^+n$ cross sections, the
SAID solution at higher energies is now far more reliable than in
previously published analyses.  Based on the earlier SAID SM05
solution, the authors of Ref. ~\cite{GDH} previously noted how well
the single-pion component of the Gerasimov-Drell-Hearn (GDH) sum rule 
integrand reproduced the
full result (including multi-pion and other-meson production). In
Fig.~\ref{fig:g4}, we extend this same comparison significantly beyond
the 2~GeV range of the SM05 solution. As seen in the figure, the FA08
solution now agrees well with the MAID07 result, but extends that
result to much higher $E_\gamma$. General agreement with the existing
GDH data \cite{Dutz} is good.

\begin{figure}[th]
\includegraphics[height=0.4\textwidth, angle=90]{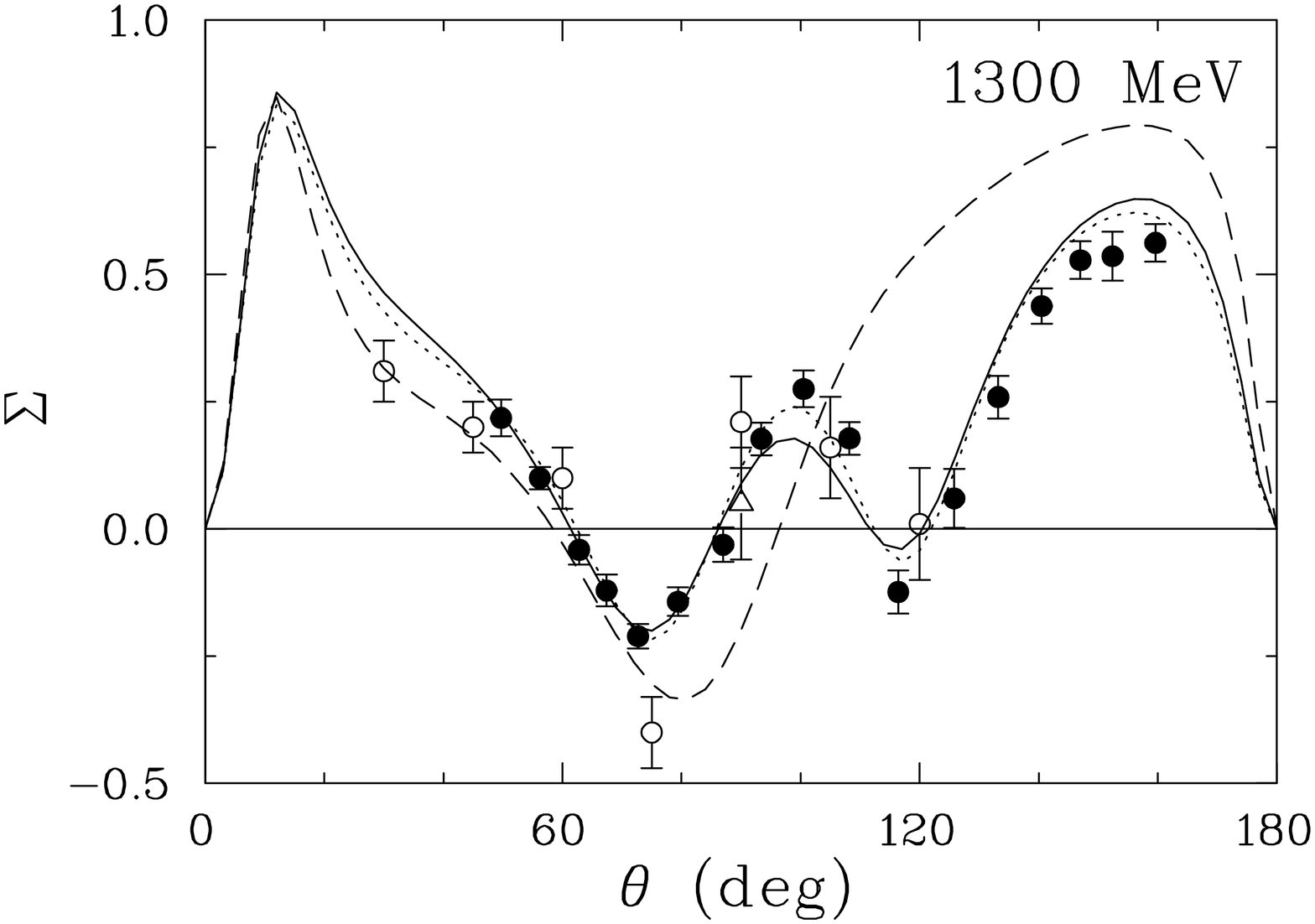}\\
\includegraphics[height=0.4\textwidth, angle=90]{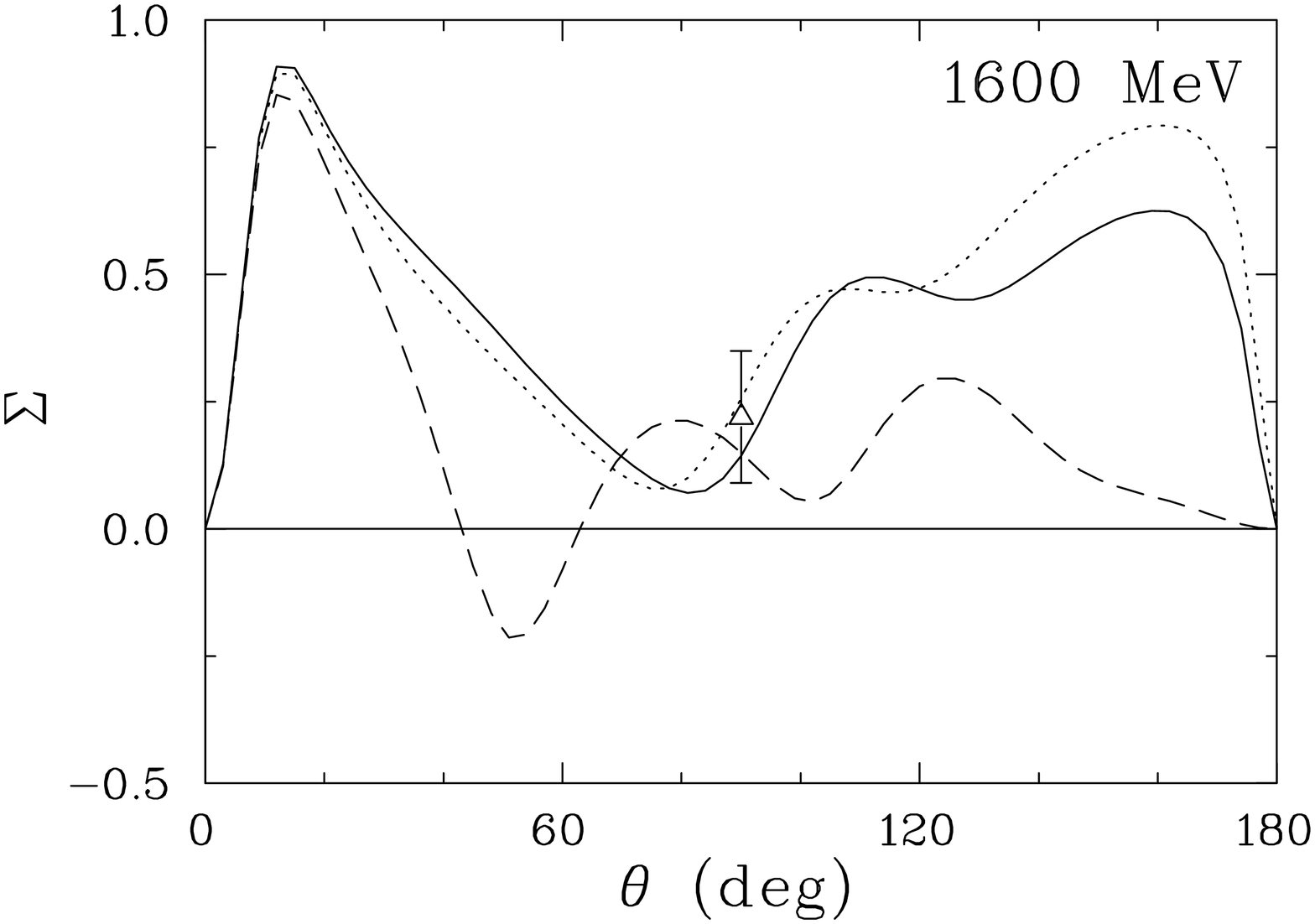}\\
\includegraphics[height=0.4\textwidth, angle=90]{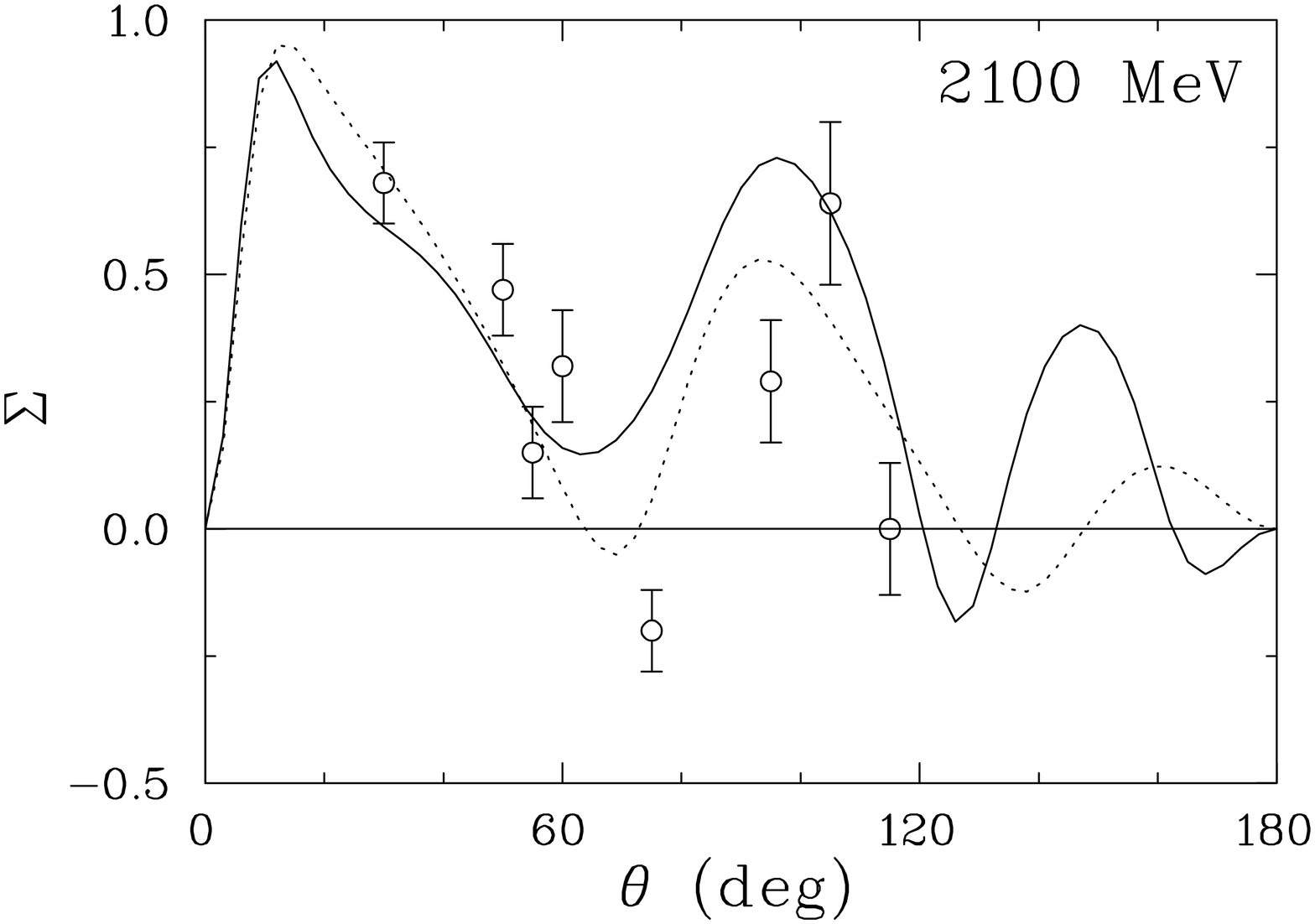}
\caption{Beam asymmetry $\Sigma$ for $\gamma p\to\pi^+n$ at
         E$_\gamma$ = 1300, 1600, and 2100~MeV vs. center-of-mass
         scattering angle.  Solid (dotted) lines correspond
         to the SAID FA08 (FA07) solution. Dashed lines give the
         MAID07 \protect\cite{Maid07} predictions.  Experimental data
         are from GRAAL (filled circles) ~\protect\cite{GRAAL}, from
         DNPL (open circles)~\protect\cite{DNPL1}, and from CEA (open
         triangles) ~\protect\cite{CEA}. Plotted uncertainties are
        statistical.  Systematic uncertainties are taken into
        account in the fit (see text). \label{fig:g5}}
\end{figure}

For completeness, we provide in Fig.~\ref{fig:g5} a comparison between
the predictions for the beam asymmetry $\Sigma$ from the FA07, MAID07,
and FA08 analyses and the experimental data for that variable from
GRAAL \cite{GRAAL}, from DNPL \cite{DNPL1}, and from CEA \cite{CEA}
for the $\gamma + p \rightarrow \pi^+ n$ reaction under study here.
The agreement with the GRAAL data for $\Sigma$ at 1.3~GeV is very good
for both SAID solutions, while there are discrepancies at
center-of-mass scattering angles greater than 75$^\circ$ between those
data and the MAID07 predictions.  All three analyses are seen to match
the single $\Sigma$ data point from CEA at 1.6~GeV, and both the FA07
and FA08 analyses provide reasonably good predictions for the DNPL
data for $\Sigma$ for positive pions at 2.1~GeV \cite{DNPL1}, although
the agreement is poorer for center-of-mass scattering angles greater
than 75$^\circ$.  However, the data for $\Sigma$ remain relatively
sparse compared to the existing data for the differential cross
sections.  New data for $\Sigma$ will help firm up the experimental
situation for this energy region, and a number of experiments are
underway at Jefferson Lab to obtain such data for pions and other
mesons \cite{FROST,g8b}.
\section{Resonance Couplings}
\label{sec:ResCoupl}


As in Ref.~\cite{dugger}, we have extracted resonance couplings from 
the modified fit (FA08) using a simple resonance plus background 
assumption, a form similar to that used in the MAID analysis,
\begin{equation}
        B(W)~(1 + i T_{\pi N} ) + T_{BW}~e^{i\phi} ,
        \label{1}
\end{equation}
where $T_{\pi N}$ is the associated full pion-nucleon $T$-matrix and 
$T_{BW}$ is a Breit-Wigner parametrization of the resonance
contribution. With this model, resonance contributions have been
determined and are listed in Table~\ref{tab:tbl3}. Values for the 
resonance mass $W_R$, width $\Gamma$, and branching fraction 
$\Gamma_{\pi N} / \Gamma$ for the various resonances were taken from 
a recent SAID analysis of pion-nucleon elastic scattering data~\cite{sp06}.
These couplings were also calculated in Ref.~\cite{dugger} after the
addition of $\pi^0p$ photoproduction data reported in that reference.

The function $B(W)$ was fit to data over an energy range spanning the 
resonance position.  In the MAID determination, $B(W)$ was given by 
the Born term. Differences between the couplings quoted here and in 
MAID therefore reflect both the impact of the present data set and a 
model-dependent uncertainty associated with the resonance extraction 
procedure.

Results based on a fit not including the present data set, presented 
in Ref.~\cite{dugger}, generally fall within one to three standard 
deviations of the present values. This stability is to be expected; 
larger deviations may occur with the addition of forthcoming polarization 
measurements.  

However, the range of couplings given in Table
\ref{tab:tbl3} requires further comment.
The two resonances coupled to a $\pi N$ $S_{11}$ state are given very
different estimates in the present analysis than those provided by 
the 2007 MAID fit and
the PDG. The PDG range for the $N$(1535) accounts for the large 
discrepancy that once existed between determinations based on $\pi N$ 
and $\eta N$ photoproduction fits. Whereas the present $\pi N$ estimate, 
the PDG central value, and older $\eta N$ photoproduction analyses 
agree on a value close to 100~GeV$^{-1.2} \times 10^{-3}$, the MAID 
2007 value has now dropped to a value consistent with the 1996 SAID 
value~\cite{said96}. This low value was criticized in a number of 
papers analyzing $\eta N$ photoproduction data measured at MAMI-B in 
Mainz~\cite{krusche}.

From the plots in Figs.~\ref{fig:g6} and ~\ref{fig:g7}, a significant 
difference between the SAID and MAID fits exists in multipoles coupled 
to the $\pi N$ $S_{11}$ and $D_{13}$ resonances. This, combined with 
differences in the assumed background contribution, likely accounts for 
the variations seen in Table~\ref{tab:tbl3}. Differences in the $N$(1650) 
couplings are largely due to difficulties in separating two nearby 
resonances in a single multipole. The present $N$(1650) photo-decay 
amplitude is consistent with that found in Ref.~\cite{dugger}, given 
the large errors. The statistical significance of any inconsistencies
with the MAID analysis cannot be 
determined, as they have not presented any uncertainties for their estimates.

Both the SAID and MAID values for the $N$(1720) coupling are very different 
from the Particle Data Group (PDG) average. The PDG range does not even 
determine a sign for this coupling. As this state has the lowest $\pi N$ 
branching fraction listed in Table~\ref{tab:tbl3}, a better determination 
may require a more favorable reaction or additional information on spin observables. 
Finally, we note that while the 
present SAID fit, the fit in Ref.~\cite{dugger}, and the PDG estimate 
for the $\Delta (1700)$ photo-decay amplitudes have remained relatively 
stable, the MAID 2007 value for $A_{1/2}$ amplitude has nearly doubled 
the MAID 2003 result. This change has resulted in both the helicity 1/2 and 3/2 
couplings being more than double the PDG estimate.

\begin{table*}[ht]
\caption{Resonance parameters for $N^\ast$ and $\Delta^\ast$
         from the SAID fit to the $\pi N$ data ~\protect\cite{sp06},
         helicity amplitudes $A_{1/2}$ and
         $A_{3/2}$ from the FA08 solution, 
         MAID07 determination~\protect\cite{Maid07},
         and average values from Ref.~\protect\cite{PDG}.
         \label{tab:tbl3}}
\vspace{2mm}
\begin{tabular}{|c|c|c|c|c|c|c|c|c|c|}
\colrule
Resonance        & 
\multicolumn{3}{c|}{$\pi N$ SAID} &
\multicolumn{3}{c|}{$A_{1/2}$ (GeV$^{-1/2}\times10^{-3}$)} &
\multicolumn{3}{c|}{$A_{3/2}$ (GeV$^{-1/2}\times10^{-3}$)} \\
\cline{2-10}
& $W_{R}$ (MeV)& $\Gamma$ (MeV) & $\Gamma _{\pi}/\Gamma$ & FA08 & MAID07 & PDG & FA08 & MAID07 & PDG \\
\colrule
$N(1535)S_{11}$      & 1547 & 188 & 0.36 &   100.9$\pm$3.0 &    66 &    90$\pm$30 & & &\\ 
$N(1650)S_{11}$      & 1635 & 115 & 1.00 &     9.0$\pm$9.1 &    33 &    53$\pm$16 & & &\\
$N(1440)P_{11}$      & 1485 & 284 & 0.79 & $-$56.4$\pm$1.7 & $-$61 & $-$65$\pm$4  & & &\\
$N(1720)P_{13}$      & 1764 & 210 & 0.09 &    90.5$\pm$3.3 &    73 &    18$\pm$30 & $-$36.0$\pm$3.9 & $-$11 & $-$19$\pm$20 \\
$N(1520)D_{13}$      & 1515 & 104 & 0.63 &   $-$26$\pm$1.5 & $-$27 & $-$24$\pm$9  &   141.2$\pm$1.7 &   161 &   166$\pm$5 \\
$N(1675)D_{15}$      & 1674 & 147 & 0.39 &    14.9$\pm$2.1 &    15 &    19$\pm$8  &    18.4$\pm$2.1 &    22 &    15$\pm$9 \\
$N(1680)F_{15}$      & 1680 & 128 & 0.70 & $-$17.6$\pm$1.5 & $-$25 & $-$15$\pm$6  &   134.2$\pm$1.6 &   134 &   133$\pm$12 \\
\colrule
$\Delta(1620)S_{31}$ & 1615 & 147 & 0.32 &    47.2$\pm$2.3 &    66 &    27$\pm$11 & & &\\
$\Delta(1232)P_{33}$ & 1233 & 119 & 1.00 &$-$139.6$\pm$1.8 & $-$140& $-$135$\pm$6 & $-$258.9$\pm$2.3& $-$265& $-$250$\pm$8 \\
$\Delta(1700)D_{33}$ & 1695 & 376 & 0.16 &   118.3$\pm$3.3 &   226 &    104$\pm$15&    110.0$\pm$3.5&   210 &     85$\pm$22 \\
$\Delta(1905)F_{35}$ & 1858 & 321 & 0.12 &    11.4$\pm$8.0 &    18 &     26$\pm$11&  $-$51.0$\pm$8.0& $-$28 &  $-$45$\pm$20 \\
$\Delta(1950)F_{37}$ & 1921 & 271 & 0.47 & $-$71.5$\pm$1.8 & $-$94 &  $-$76$\pm$12&  $-$94.7$\pm$1.8& $-$121&  $-$97$\pm$10 \\ 
\colrule
\end{tabular}
\end{table*}

\section{Conclusion}
\label{sec:conc}


Differential cross sections for $\pi^+$ meson photoproduction on the
proton via the reaction $\piplusRxn$ have been determined with a
tagged-photon beam for incident photon energies from 0.725 to
2.875~GeV. All derived cross sections were based on a $\pi^+n$ missing
mass reconstruction. The relative cross sections were determined from
yields derived from a peak isolated above a well-determined
background, using Monte Carlo simulations to determine the $\pi^+$
acceptance in the CLAS spectrometer.  The relative differential cross
sections were converted to absolute differential cross sections by
measurements of the incident photon flux.

These data have been included in a SAID multipole analysis, resulting
in a new SAID solution, FA08. Comparisons to earlier SAID fits and a
fit from the Mainz group show that the new solution is much more
satisfactory at higher energies.  Although resonance couplings have not
changed significantly with the addition of these cross sections to the
world data set, significant changes have occurred in the high-energy
behavior of the SAID cross-section predictions and amplitudes, as can
be seen in Fig.~\ref{fig:g2} for the cross-section and
Fig.~\ref{fig:g4}, for the single-pion contribution to the GDH sum
rule.  Further improvement will be possible with future measurements
of spin observables for the photoproduction process that can be
expected from FROST~\cite{FROST} and the {\tt{g8b}} CLAS running
period~\cite{g8b}.

\clearpage
\acknowledgments

The authors gratefully acknowledge the
work of the Jefferson Lab Accelerator
Division staff.  This work was supported by the National Science
Foundation, the U.S. Department of Energy (DOE), the
French Centre National de la Recherche Scientifique and Commissariat
\`a l'Energie Atomique, the Italian Istituto Nazionale di Fisica
Nucleare, and the Korean Science and Engineering Foundation.  The
Southeastern Universities Research Association (SURA) operated
Jefferson Lab for DOE under contract DE-AC05-84ER40150 during this
work.

\clearpage



\end{document}